\documentclass[sigconf]{acmart}

\AtBeginDocument{%
  \providecommand\BibTeX{{%
    \normalfont B\kern-0.5em{\scshape i\kern-0.25em b}\kern-0.8em\TeX}}}

\setcopyright{acmcopyright}
\copyrightyear{2018}
\acmYear{2018}
\acmDOI{10.1145/1122445.1122456}

\copyrightyear{2021}
\acmYear{2021}
\setcopyright{acmlicensed}\acmConference[CHI '21]{CHI Conference on Human Factors in Computing Systems}{May 8--13, 2021}{Yokohama, Japan}
\acmBooktitle{CHI Conference on Human Factors in Computing Systems (CHI '21), May 8--13, 2021, Yokohama, Japan}
\acmPrice{15.00}
\acmDOI{10.1145/3411764.3445195}
\acmISBN{978-1-4503-8096-6/21/05}

\title[Vis Ex Machina]{Vis Ex Machina: An Analysis of Trust in Human versus Algorithmically Generated Visualization Recommendations}

\author{Rachael Zehrung}
\authornote{Both authors contributed equally to this research.}
\email{rzehrung@umd.edu}
\affiliation{\institution{University of Maryland}}

\author{Astha Singhal}
\authornotemark[1]
\email{asingha3@terpmail.umd.edu}
\affiliation{\institution{University of Maryland}}

\author{Michael Correll}
\email{mcorrell@tableau.com}
\affiliation{\institution{Tableau Research}}

\author{Leilani Battle}
\email{leibatt@umd.edu}
\affiliation{\institution{University of Maryland}}

\usepackage[normalem]{ulem}
\usepackage{subcaption}
\usepackage{caption}

\usepackage{enumitem}
\usepackage{xcolor}
\newif\ifnotes
\notestrue

\definecolor{highlight}{RGB}{0,0,0}

\newcommand{\highlight}[1]{\textcolor{highlight}{#1}}

\renewcommand{\shortauthors}{Zehrung and Singhal et al.}


\begin{teaserfigure}
    \centering
  \includegraphics[width=0.9\textwidth]{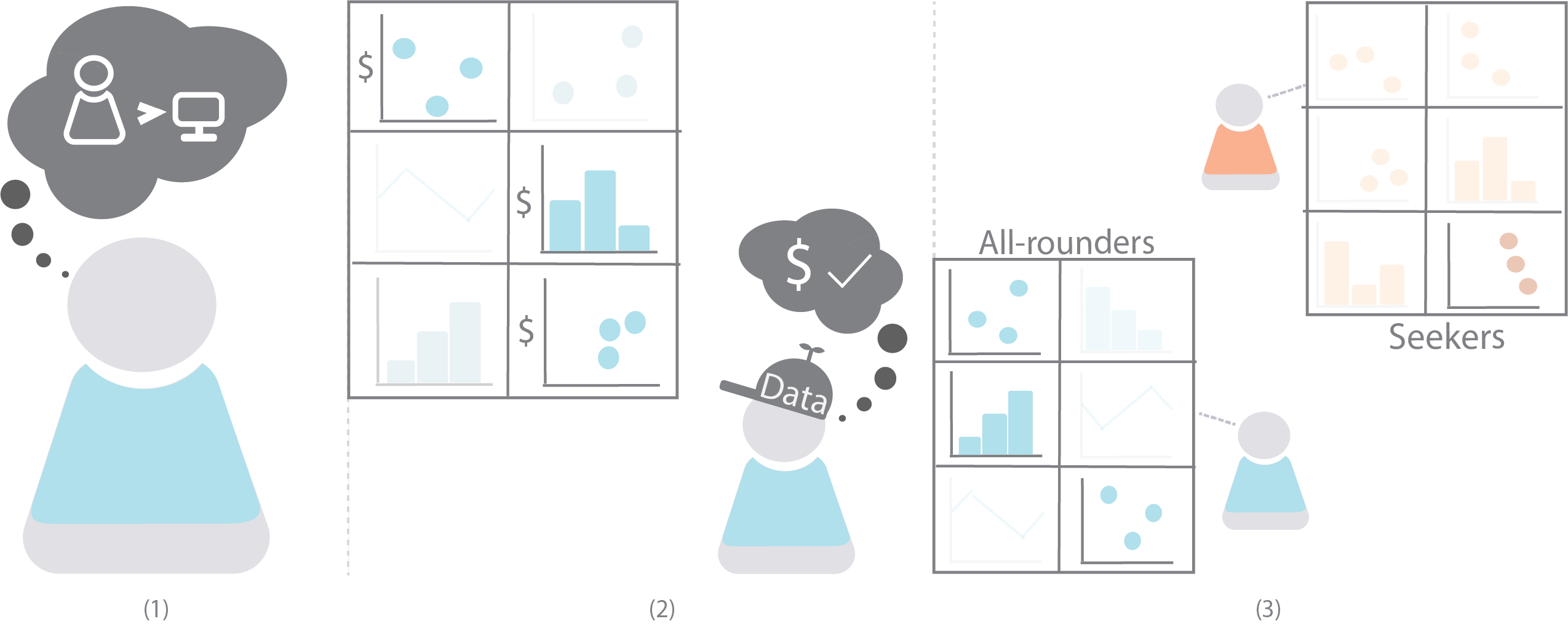}
  \caption{What factors influence why and how much people trust visualization recommendations? We present results from an exploratory study of how people interpret visualization recommendations from different sources (human or algorithm). We find that (1) participants generally express an a priori preference for recommendations provided by humans, but (2) seem to evaluate recommendations based on the inclusion of data attributes they find most relevant. Our results (3) point to the existence of differing patterns of information foraging among viewers, who seem to be largely unaffected by source. }
  \Description{Subfigure 1 has an icon of a person thinking that humans are greater than computers, expressing that they have a preference for human recommendations. Subfigure 2 shows an icon wearing a "data" thinking cap, and examining a panel of 6 visualizations. 3 of the visualizations contain a "dollar sign" attribute, and the icon likes that it contains those attributes. Subfigure 3 has 2 icons looking at 2 different panels. One of the panels is labelled all-rounders and the other is labelled seeker. The all-rounder panel has 3 emphasized graphs, and the seeker has only 1 emphasized graph. }
  \label{fig:teaser}
\end{teaserfigure}

\begin{document}
\begin{abstract}

More visualization systems are simplifying the data analysis process by automatically suggesting relevant visualizations. However, little work has been done to understand if users trust these automated recommendations. In this paper, we present the results of a crowd-sourced study exploring preferences and perceived quality of recommendations that have been positioned as either human-curated or algorithmically generated. We observe that while participants initially prefer human recommenders, their actions suggest an indifference for recommendation source when evaluating visualization recommendations. The relevance of presented information (e.g., the presence of certain data fields) was the most critical factor, followed by a belief in the recommender's ability to create accurate visualizations. Our findings suggest a general indifference towards the provenance of recommendations, and point to idiosyncratic definitions of visualization quality and trustworthiness that may not be captured by simple measures. We suggest that recommendation systems should be tailored to the information-foraging strategies of specific users.

\end{abstract}



\maketitle
\section{Introduction}
\label{sec:introduction}


As a field and industry, visual analytics is beginning to incorporate increasingly automated methods into its processes~\cite{heer2019agency}, often to create visualization recommendations. From academia, visualization systems like Draco~\cite{moritz2018formalizing}, Data2Vis~\cite{dibia2018data2vis}, and Tableau's Show Me Feature~\cite{mackinlay2007show} attempt to automatically generate expressive and informative visualizations from a dataset. In industry, features like PowerBI's ``Quick Insights'' panel~\cite{powerbi} attempt to present quick visual summaries of interesting or important aspects of a dataset. 


While the relative trustworthiness of machine learning models has been investigated in other contexts~\cite{castelo2019let,jakesch2019ai,Yin2019accuracytrust} there has been little investigation of viewers' trust of recommendations in visual analytics. Though widely used visualization authoring software like Tableau and PowerBI incorporate recommendations, users may adopt the authoring portion of the tool without trusting or utilizing the recommendation features; many recommendation systems in popular tools such as ``Explain Data'' in Tableau \cite{tableauexplain} are relatively new, with unclear adoption. If analysts are wholly trusting of algorithmic recommendations, potentially biased or inaccurate results could result in poor decision-making \cite{correllEthical}. However, if analysts habitually devalue automated insights or recommendations, existing research efforts into automated visualization recommendations may be misaligned with their needs. 

In this paper, we present the results of a pre-registered, exploratory human-subjects study on how the perceived source (human or algorithmic) of visualization recommendations impacts assessments of the utility of those recommendations for general audiences. We sought to determine if existing attitudes and biases regarding algorithmic recommendations on the whole would impact people’s assessments of the quality of visualization recommendations, and whether these biases would persist even as we adjusted the anticipated relevance of the recommendations.


Through a quantitative and qualitative analysis of our collected data \footnote{Our study materials, including data tables and analyses scripts, are available at \url{https://osf.io/zmnh3/?view\_only=c3a9a1568d554c3587132b339c72f22e}},
we found that participants initially preferred human-curated recommendations, but tended to be source-agnostic when evaluating visualization recommendations of equal quality. This appeared to hold even across different levels of analytics experience. Participants' evaluations of recommendation sources seemed to emphasize the degree of overlap between the participant's top attributes of interest and the attributes displayed in the recommendations. In stating their rationale for preferring one set of recommendations over another, participants fell into two categories of behavior: \emph{all-rounders} tended to focus on the quality of recommendations as a whole, while \emph{seekers} honed in on the presence of particular visualizations or attributes.

Our findings partially support existing assumptions in the community that users trust automated visualization recommendation systems. Though some participants held onto folk theories about the capabilities of a given recommendation source, users on the whole exhibited different mental models on evaluating the utility of recommendation panels. These observations suggest that users are not uniform in how they evaluate, and subsequently determine the utility of, visualization recommendations.
We reflect on how designers can present recommendations to a broad range of users in a way that mitigates the risk of user bias in interpreting the results, contributing to an emerging body of work on algorithmic trust.
\section{Related Work}
\label{sec:related-work}
Our research questions and experiment design are informed by existing assumptions for visualization recommendation systems, as well as studies measuring users’ preferences for algorithmically generated recommendations in other contexts. We highlight three topics of related research: visualization recommendation systems, inclusion of contextual information, and trust in algorithmic decision making.

\subsection{Visualization Recommendations}
\label{sec:related-work:recommendations}
A number of systems recommend sets of visualizations based on various assumptions (or explicit solicitation) of information and patterns that users would find valuable. Golfarelli et al.~\cite{golfarelli2019model} propose a pipeline for generating visualization recommendations based on a set of \highlight{predefined} user objectives. Wongsuphasawat et al.~\cite{wongsuphasawat2015voyager,wongsuphasawat2017voyager} provide flexibility by allowing users to specify partial visualization designs, and recommend visualizations that extend these partial specifications. 
Vartak et al. identify statistically significant differences between sub-populations within a dataset, and recommend bar charts capturing these differences~\cite{vartak2015s}. 

The number of recommendations vary per system as well, which impacted our experimental design. Some systems, such as Calliope~\cite{calliope_2020} and ``Retrieve Then Adapt''~\cite{retrievethenadapt}, focus on recommending a single visualization, such as a data story or infographic. In most cases, users are provided an ensemble of recommendations grouped together in a single \emph{panel}. Voyager is a salient example, and so we adopted a similar design for our study~\cite{wongsuphasawat2015voyager,wongsuphasawat2017voyager}.

Though these systems employ differing strategies, they appear to be developed under the implicit assumption that users generally want and trust algorithmically-generated visualization recommendations. Given that the differences between human and algorithmically-generated visualizations are not well understood, we shed light on user trust in recommendation source through the use of labelling, similar to Jakesch et al. and Shank et al. ~\cite{jakesch2019ai, shank2013computers}. That is, we present participants with visualizations that were all created by humans, but some were labelled otherwise.

\subsection{Inclusion of Contextual Information}
\label{sec:related-work:metadata-pref}

Several projects have explored the potential ways in which additional contextual information impacts how people interpret and evaluate visualizations.

The ``Contestability in Algorithmic Systems'' Workshop at CSCW 2019 argues for the inclusion of humans in the loop of algorithmic decision making, especially as decisions made by machine learning systems have greater consequence. One of their design objectives to do so is through \textit{legibility}, in which systems would include explanations for the decisions made and conclusions drawn \cite{contestability}. 

There are many projects that explore legibility, especially in machine learning systems. Yang et al.~\cite{YangVisualExplanations} found that additional metadata, in the form of example-based explanations for machine-learning classifiers, did improve user trust, although Kizilcec~\cite{Kizilcec2016} \highlight{found that} that providing \textit{too much} explanatory information can erode trust. Cheng et al.~\cite{cheng2020dece} take this one step further and propose DECE, a visual interactive system to better understand the decision rules of machine learning systems through counterfactual explanations. 

However, these systems are focused on machine learning in particular. Peck et al. explore the notion of metadata, in this case the source of the data, in relation to perceptions of visualizations~\cite{peck2019data}. Through semi-structured interviews with residents of rural Pennsylvania, they \highlight{gathered} initial perceptions and rankings of visualizations without participants knowing the sources of visualizations. Afterwards, they \highlight{revealed} the sources and \highlight{asked} whether knowing the sources of the graphs impacted how participants viewed them, as well as their overall ranking. They found that 60\% of participants chose not to re-rank their visualizations after revealing the sources, indicating that for some, additional context about source may not impact the perceived utility or credibility of a visualization.

We seek to understand the impact of additional contextual information on users’ visualization preferences. However, we study this topic from the higher-level perspective of gauging trust in particular \emph{sources} of recommendations (i.e., human or algorithm).

\vspace{-0.5em}
\subsection{Algorithmic Decisions and Trust}
\label{sec:related-work:decisions-trust}

Several projects investigate how people respond to recommendations or decisions made by algorithms.
Victor et al.~\cite{victor2011trust} explain how users tend to trust recommendations from known entities (particularly people) more than unknown entities (i.e., complete strangers or algorithms); 
Lee~\cite{lee2018understanding} finds that users seem to distrust managerial decisions made by algorithms, due in part to a feeling of dehumanization by algorithms and a lack of shared social understanding. 
In contrast to the above studies that showcase negative responses to algorithmic decision-making, Logg et al.~\cite{logg} found that people were more likely to adhere to advice when they \highlight{believed} it was given by an algorithm than by a person. 
In situations where algorithmic performance is ambiguous (i.e., neither clearly good nor clearly poor), users' generalized implicit attitudes towards automation impact their propensity to trust a specific automated system~\cite {merrittAmbiguous}. These studies indicate that user trust in algorithmic decisions appears to be very situational in nature.


A number of projects evaluate people's perception of interactions with agents declared to be algorithmic or human, when in fact, the source is held constant. 
Shank~\cite{shank2013computers} finds that people perceive ``organizations as more responsible and in control when they have employed human, not computer, representatives.'' Jakesch et al.~\cite{jakesch2019ai} find that people seem to trust renters on \highlight{Airbnb} more when they write their own profiles compared to renters whose profiles \highlight{they believe} are generated by AI; however, this effect is only observed when human-written and AI-generated profiles are compared side-by-side. Graefe et al.~\cite{2016newscredibility} go one step further by modifying both the declared and \textit{actual} source of news articles (computer or human written). They find that while modifying the declared source had small but consistent effects in favor of human-written articles, modifying the actual source had larger effects. Participants generally regarded computer-written articles as more credible but less readable. 

We see many observations in the literature of users preferring interactions with, recommendations from, and unilateral decisions by humans over algorithms, with some exceptions (e.g., \cite{edwards2014bot}). However, no existing studies explicitly measure user trust in sources of recommendations for visual analytics, so it is unclear to what degree these results also apply to visualization. In this paper, we present a first step towards measuring user trust in human-curated versus algorithm-generated visualization recommendations.


\section{Motivation and Research Questions}
\label{sec:motivation}
When we compare the literature on user trust in algorithmic decision-making to that of visualization recommendations, we find a contradiction: visualization recommendation features are designed as if users will naturally trust them, yet in other contexts users express a clear distrust of algorithmic decision-making. To the best of our knowledge, no existing studies explicitly measure user trust in algorithm-generated (versus human-curated) visualization recommendations. 

The absence of research on the perceived trustworthiness of existing recommendation systems, as well as the potential mismatch between these systems and human mental models, serve as the core motivation for our work:
if the user is biased against recommenders, then expending resources on generating recommendations may prove wasteful, but if the user blindly trusts all recommendations (including occasional bad ones), the user may inadvertently draw inaccurate (and potentially dangerous~\cite{bresciani2015pitfalls}) conclusions. 
In this work, we seek to explore the following research question: 

\noindent\textbf{Research Question 1}: \emph{How do existing preferences for human-curated versus algorithmically-generated recommendations affect the evaluation of recommendation quality or utility?}

Specifically, we seek to understand whether the source of recommendations (human or algorithmic) may alter a user's perception of the recommended charts. Obtaining a deeper understanding of why users prefer certain recommendations over others can help visualization system designers to better employ techniques to support a wide range of users. To further understand the context for user preferences, we explore the following secondary analysis questions:

\noindent\textbf{Research Question 2}: \emph{If clear preferences are observed, what reasons do users give for preferring certain recommenders?}


\noindent\textbf{Research Question 3}: \emph{What effect (if any) does statistical or data analysis experience have on user preferences?}




\noindent\textbf{Exploratory Hypotheses}:
Based on the above research questions, and to capture our expectations for how user trust may (or may not) manifest in our study, we formulated two exploratory hypotheses:

\begin{itemize}[nosep]
    \item \textit{People will rate recommenders based on how much the recommendations overlap with preferred data attributes.} 
    \item \textit{Prior preferences for human or algorithmic recommendations will bias participants towards the corresponding panel, even in the face of differing amounts of recommendation relevance.} 
\end{itemize}
\section{Experimental Design}
\label{sec:experiment-design}
\begin{table*}[tb]
\begin{tabular}{l l}
\hline
\textbf{Attribute Category} & \textbf{Attribute Names}                              \\ \hline
\textbf{Ratings}            & \begin{tabular}[c]{@{}l@{}}IMDB Vote Average (Q), IMDB Vote Count (Q)\end{tabular}                                                                    \\ \hline
\textbf{Finances}           & \begin{tabular}[c]{@{}l@{}}Budget (Q), Domestic Gross (Q), Profit (Q), Worldwide Gross (Q)\end{tabular}                                                \\ \hline
\textbf{Details}            & \begin{tabular}[c]{@{}l@{}}MPAA rating (O), Country (N), Genre (N), Release Date (T), Runtime (Q)\end{tabular}                                      \\ \hline
\textbf{Popularity}         & \begin{tabular}[c]{@{}l@{}}Facebook Likes by Cast (Q), Facebook Likes by Lead Actor (Q),\\ Facebook Likes by Movie (Q), Popularity (Q)\end{tabular} \\ \hline
\end{tabular}
\caption{All attributes evaluated in our merged Movies dataset, grouped by attribute category. Data types are specified in parentheses: \textbf{Q}uantitative, \textbf{O}rdinal, \textbf{N}ominal, or \textbf{T}emporal.}
\label{tab:attributes} 
\Description{There are four attribute categories: ratings, finances, details, and popularity. Ratings includes the IMDB Vote Average (Q) and the IMDB Vote Count (Q). Finances includes Budget (Q), DOmestic Gross (Q), Profit (Q), and Worldwide Gross (Q). Details includes MPAA Rating ()), Country (N), Genre (N), Release Date (T), and Runtime (Q). Popularity includes Facebook Likes by Cast (Q), Facebook Likes by Lead Actor (Q), Facebook Likes by Movie (Q), and Popularity (Q).}
\vspace{-5mm}
\end{table*}
We designed \highlight{an} online experiment to explore the relationships between visualization recommendations, user preference and trust. \footnote{Our research questions and exploratory hypotheses were pre-registered before running the experiment, available at \url{https://aspredicted.org/blind.php?x=tu5jk3}} In this section, we detail the design of our experiment, as well its limitations and trade-offs.

\subsection{Participants}
\label{sec:experiment-design:participants}

We recruited 114 participants via \url{Prolific.ac}, a crowdsourcing platform comparable to Amazon Mechanical Turk~\cite{peer2017beyond}. They were compensated \$3.35 for completing the experiment with an estimated completion time of 20 minutes, resulting in a projected \$10.05 hourly wage (and actual average wage of \$12.57 per hour). 

Recruited participants were at least 18 years of age with baseline data analysis experience (e.g., having taken a data science course, or worked in analytics). We chose to not impose more restrictive filters on experience to allow for a more diverse participant pool.
77\% of our participants identified as female. 80.5\% of participants were 18-24 years of age. 66.4\% had at least some college education.  

\vspace{-0.5mm}
\subsection{Experiment Dataset}
\label{sec:experiment-design:dataset}

In keeping with our broad recruitment criteria, we used movies as the basis for our experimental task because it is a domain with which the general public is somewhat familiar, and has extensive publicly available data.
We pooled attributes from three  movies datasets from Kaggle \cite{kaggle1,kaggle2,kaggle3} and the-numbers.com \cite{thenumbers} to provide a diversity of data attributes to explore. Certain attributes were omitted from our analysis for the following reasons:
\begin{itemize}[nosep]
    \item the attribute was redundant with another, already selected attribute
    \item the distribution of values was highly skewed (e.g., language)
    \item the attribute caused excessive visual clutter due to high cardinality (e.g., title, director)
    \item the attribute contained multiple values per tuple
\end{itemize}

The attributes list (\autoref{tab:attributes}) was provided to participants to give them an idea and explanation of what attributes could be explored, and was later used to create visualization recommendations.

The dataset was cleaned to prevent visualizations from having significant occlusion, extreme outliers, or other distracting artifacts. First, we sampled 200 tuples. Then for each visualization, we filtered out any outliers (i.e., tuples outside of the interquartile range) for each attribute rendered in the visualization. As our study examines the relevance of a visualization in terms of attributes, bivariate visualizations were generated that employ standard best practices for attribute encodings. Quantitative and temporal data is encoded using position, and categorical data with length \cite{mackinlay2007show}.


\subsection{Experiment Overview}
\label{sec:experiment-design:overview}

\begin{figure*}[tb]
\centering
\includegraphics[width=0.7\textwidth]{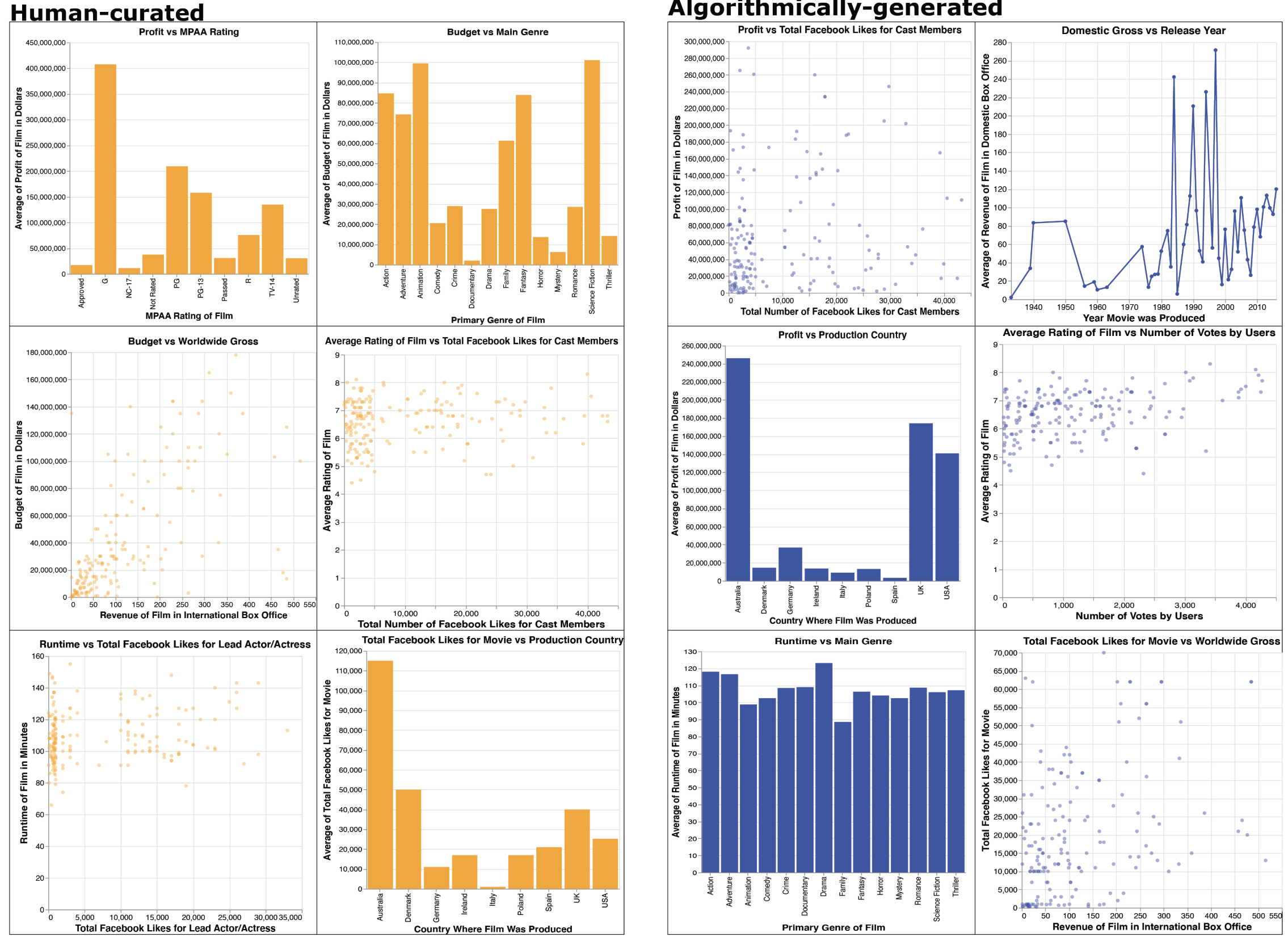}
\caption{A snapshot of two visualization recommendation panels from the experiment interface. Each panel is labeled as either ``Human-Curated'' or ``Algorithmically-Generated,'' and assigned a random order position (left/right) and color (orange/blue). Color was used to further distinguish the recommendation source.} 
\label{fig:panels-example}
\Description{2 panels of 6 visualizations each are placed side by side, with the human curated panel on the left, and algorithmically curated panel on the right. Each panel is comprised of bivariate graphs, and are either bar graphs, scatter plots, or line graphs.}
\end{figure*}


Participants were given time to review and complete the consent form before beginning the study, and could withdraw at any point. Participants took 17 minutes on average (s.d. 9 minutes) to complete the survey. Those that were 3 standard deviations below the average time were to be excluded from our analysis, though none fell into this category.

We then gave participants the following scenario: ``a well-known film studio is hosting a competition for new movie ideas, and you are currently gathering information in preparation for a pitch on why your movie would be successful. To help you extract insights about successful movies, the studio has provided you with various metrics about movies that have been created in the past.''


Our study progressed in three phases, where participants:
\begin{itemize}[nosep]
    \itemsep0em
    \item \textbf{Phase 1}: record recommendation source preferences and attribute preferences for a given movies dataset;
    \item \textbf{Phase 2}:  rate two separate groups (or panels) of visualization recommendations, and select one panel to proceed with; and
    \item \textbf{Phase 3}:  complete surveys collecting experiences with the recommendation panels and demographic information.
\end{itemize}
A pilot study of 12 participants was run prior to our main study to test our materials and procedure. 

\subsection{Phase 1: Recording Prior Preferences}
We \highlight{captured} participants' priors in two ways: by asking participants to rank the movies attributes they \highlight{were} most interested in analyzing, and by asking participants about their general prior preferences for human-curated versus algorithm-generated recommendations.

\subsubsection{Ranking Attributes of Interest}

Before seeing any visualization recommendations, participants were provided a list of all data attributes (with corresponding explanations) present in the movies dataset, and asked to rank the top five attributes they would like to reference while creating their pitch.

This ranking task helped participants to familiarize themselves with the available attributes for analysis, and helped us to gain an initial understanding of the attributes participants wanted to see in later visualization recommendation panels. While these rankings were not used to generate visualizations, they helped us assess the expected relevance and utility of the panels that were displayed to participants in the later stages of the experiment.

\subsubsection{Soliciting Prior Recommendation Preferences} 
We then asked participants whether they prefer recommendations from humans, algorithms, or neither (i.e., no preference), in the context of song recommendations. Though song recommendations may not fully generalize to the context of visualization recommendations, we aimed to capture prior preferences in a familiar context where many people encounter both automated (e.g., Spotify mixes~\cite{spotify}) and human recommendations (e.g., personally curated playlists). This mixture of familiarity and transferability was useful for the task without narrowing the participant pool to those with direct experience with visualization recommendation systems.


\subsection{Phase 2: Comparing Recommendation Panels}
\label{sec:experiment-design:comparing}

Once the participant's initial preferences were recorded, the participant was then asked to compare two separate panels of visualizations (see \autoref{fig:panels-example} for example panels).  These panels are comprised of 6 visualizations of 3 possible types: line chart, bar chart, or scatterplot. Our focus on panels over individual charts is motivated by the design of existing recommendation systems (e.g.,~\cite{wongsuphasawat2015voyager, mackinlay2007show, vartak2015s}) which present multiple visualization recommendations at a time. One panel was labeled  as ``Human-Curated'' and the other as ``Algorithmically-Generated.'' However, both were in fact generated by the authors and these labels were assigned randomly.

Participants were asked to rate how useful each panel would be for completing the task (analyzing data to support a movie pitch) on a ten point Likert scale, and to select a single panel for their analysis. After making a selection, participants were asked to explain their choice of panel. The ratings and final selection enabled us to gather information on the perceived relevance of each panel.

 
 After a specific recommendation panel was selected (either human or algorithm), the participant was asked to describe what they learned from this panel that would help them in writing a movie pitch. The participant was then asked to rate the panel based on how helpful it was in forming insights based on a 10 point Likert scale, and asked if there was anything they would have liked to see as part of the recommendations.

\subsection{Phase 3: Completing Final Surveys}
The final phase of the experiment involved completing two surveys for capturing participants' decision-making processes and overall impressions, as well as demographic information.

First, participants were asked a series of 8 questions asking the extent to which participants agreed to various statements about the quality of human recommendations and algorithm recommendations, and any perceived differences between the two recommendation sources.
For example, participants indicated their agreement with such statements as ``The algorithm did a good job in selecting the visualizations that I should analyze,'' ``I felt that key visualizations were often missing from the recommendations,'' and ``There was a noticeable difference in the quality of recommendations between humans and algorithms'' on a 5 point Likert scale. The survey concluded with demographic questions, including collecting gender, age, and educational attainment to prevent priming. 

\begin{figure*}
    \centering
    \includegraphics[width=.9\textwidth]{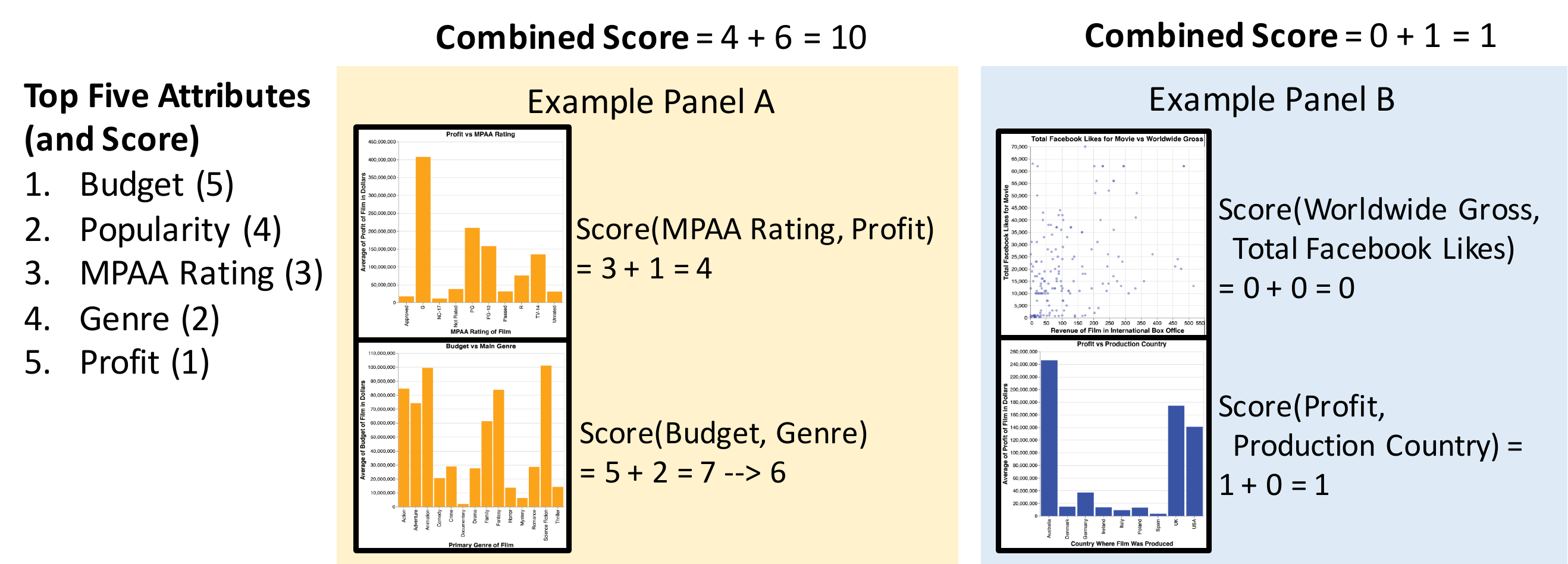}
    \caption{A demonstration of how visualization recommendation panels are scored, with four visualizations taken from \autoref{fig:panels-example} as examples. Higher ranking attributes award more points than lower-ranking attributes, and the narrow dynamic range of high-impact panels causes us to remap all panels with particularly high scores.}
    \Description{Examples of how to calculate visualization panel relevance scores based on 5 attributes (budget with a weight of 5, popularity with a weight of 4, MPAA Rating with 3, Genre with 2, and Profit with 1). 2 example panels are displayed. Panel A has a score of 10, because one visualization has a score of 6 and the other score a 4. These visualizations got their scores because one visualized MPAA Rating (3) with profit (1), which sums to 4. Similarly, the other visualization was budget(5) and Genre (2) which sums to 7. All ratings greater than 6 get mapped to 6.}
    \label{fig:scoring-example}
\end{figure*}
\vspace{-0.3mm}
\subsection{Computing Recommendation Panel Relevance}
To better understand how participants’ panel selections may be influenced by prior preferences, we designed four panels of varying relevance. In this way, we could see whether participants’ panel selections were in alignment with the perceived relevance of the panels (i.e., in alignment with their top five attributes of interest). 

\label{sec:experiment-design:relevance}

\subsubsection{Scoring Individual Visualizations}
Before scoring the quality of entire panels, we first created a method to score individual visualizations.
We had performed an initial pilot study with 12 participants, where we created panels from 5 randomly selected attributes but otherwise followed the current study design. Through that pilot, we found that the top five most popular data attributes were (in order of frequency): budget, popularity, MPAA rating, genre, and profit (also shown in \autoref{fig:scoring-example}).
These attributes served as the basis for constructing high and low relevance visualization recommendations.
We applied a simple linear weighting system to assign a weight to each of these attributes (budget received a weight of five, popularity a weight of four, and so on). All other attributes were assigned a weight of zero. Though other weighting mechanisms, such as exponential or quadratic, could have been used, as participants were ranking them in a linear fashion, we felt it prudent to use a linear weighting to match. 

At first, a single visualization ranged from a score of 0--9. As only a few visualizations could have a score of 6--9, we clamped all scores higher than 6. The resulting 0--6 per-visualization scoring lends itself to a more even distribution. Relevance for the bivariate visualization was calculated by summing the weights of the two corresponding attributes.
For example, consider the first visualization in example panel A of \autoref{fig:scoring-example}, which visualizes MPAA Rating versus Profit. MPAA Rating has a weight of 3, and Profit a weight of 1, producing a summed score of 4. In contrast, the first visualization of example panel B has no relevant attributes, producing a relevance score of 0.

%
%

\subsubsection{Panel Generation Strategy}
\label{sec:experiment-design:panel-generation}
Using the scoring mechanism for individual visualizations, we then computed scores for entire recommendation panels. For example, to calculate the final scores for example panels A and B in \autoref{fig:scoring-example}, we sum the corresponding visualizations, producing combined scores of ten and one, respectively.
Note however that each panel from our study consists of six separate visualization recommendations.
Three of these visualizations were chosen to be relevant (i.e., with a score greater than zero), and the rest were selected to be irrelevant (i.e., have a score of zero).
%
%
With possible panel scores ranging from 3--18, to compute two ``high relevance'' panels, we generated two recommendation panels with scores in the range of 13--18. To compute two ``low relevance'' panels, we generated two panels with scores in the range of 3--8. For example, panel A from \autoref{fig:scoring-example} could be the starting point for building a ``high-relevance'' panel, and panel B a ``low-relevance'' one. Panels were also chosen to have at most one repeating data attribute to ensure a breadth of visualized attributes.

\subsubsection{Permuting Panel Pairings}
\label{sec:experiment-design:pairings}

Given our two ``high relevance'' panels and two ``low relevance'' panels, we have six total pairing scenarios: high-high (one pairing), high-low (four pairings), and low-low (one pairing). Taking panel order into account, we have 12 possible ordered pairs for comparison. Each ordered pair represents a separate condition of our experiment. Color and ordering of human versus algorithm labels were controlled by randomizing across participants. We collected data until at least eight participants completed each of our 12 experiment conditions.


\subsection{Experiment Design Limitations \& Trade-offs}
\label{sec:experiment-design:limitations}

\begin{figure*}[htbp]
\centering
\begin{minipage}{0.9\columnwidth}
  \centering
\includegraphics[width=\textwidth]{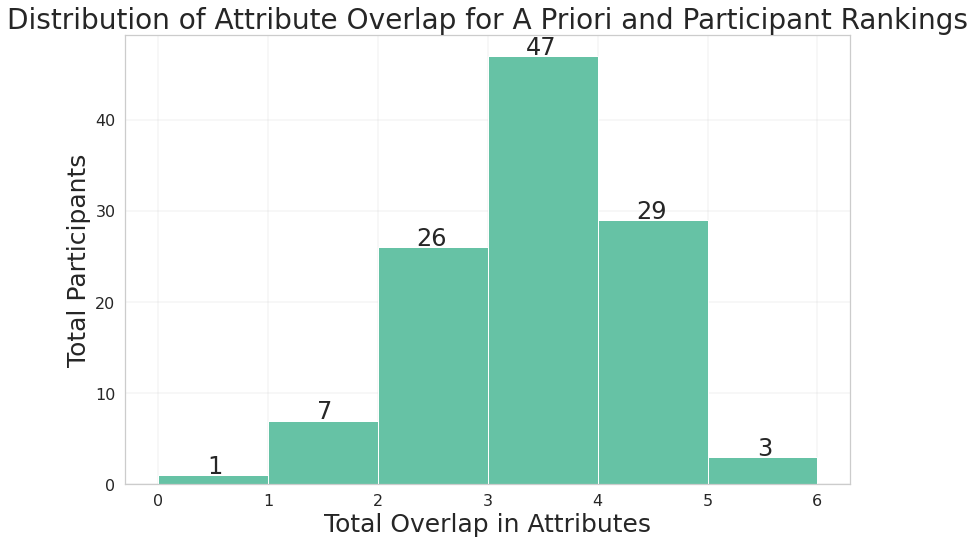}
\subcaption[first caption.]{A distribution of panel rating by number of common \\ features, or attribute overlap.}
\Description{A histogram of total attribute overlap. The histogram has a slight right skew, with most participants having at least 3 attributes overlap.}
\end{minipage}%
~%
\begin{minipage}{0.9\columnwidth}
  \centering
\includegraphics[width=\textwidth]{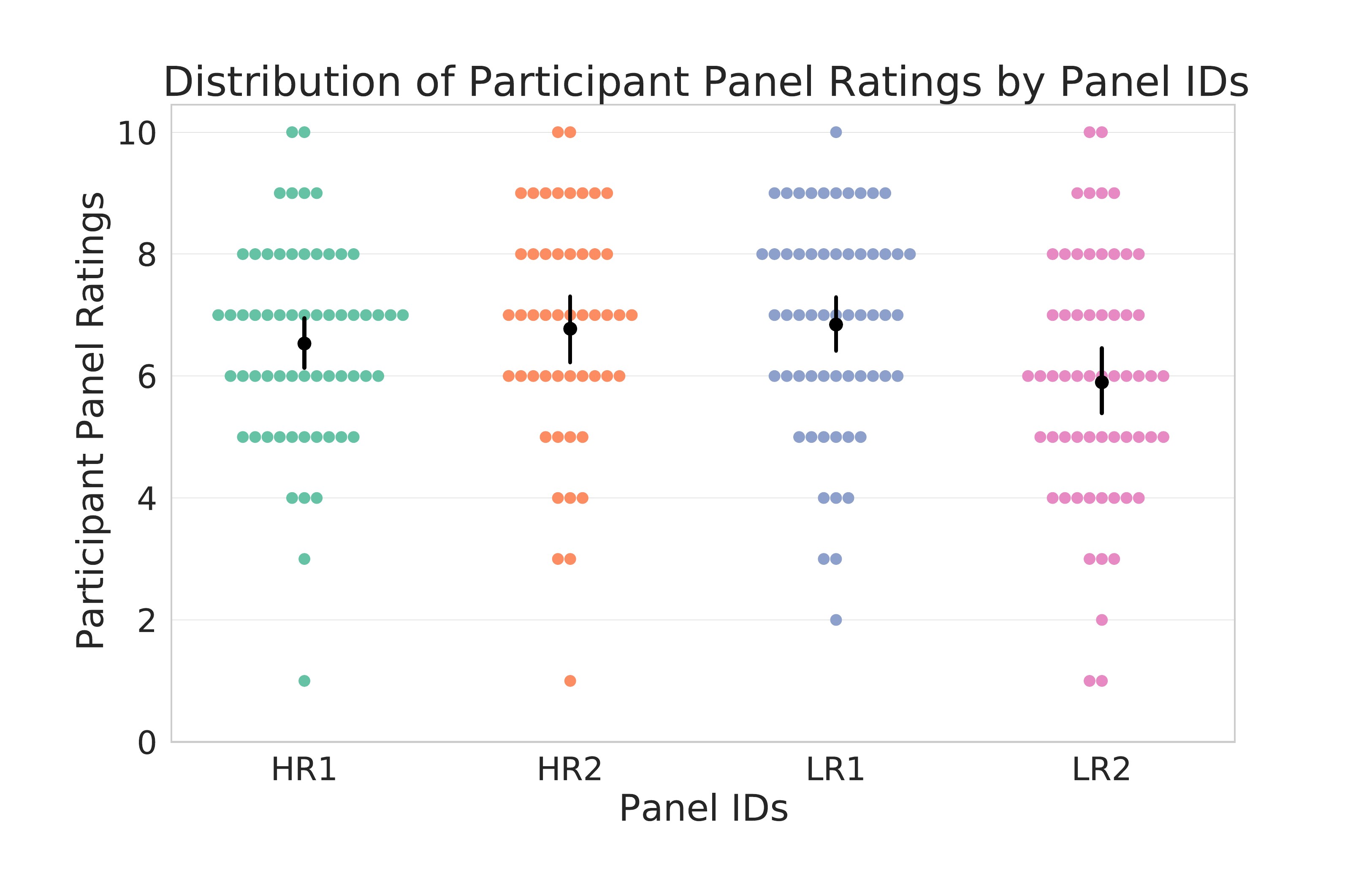}
\subcaption[second caption.]{A distribution of relevances by panel ID, calculated with a 95\% confidence interval and 1000 bootstrapped iterations. HR refers to a high relevance panel, and LR refers to a low relevance panel.}
\Description{A swarm plot visualizing the distribution of relevance scores by panel ID (HR1, HR2, LR1, and LR2). Participants rated all panels similarly, with ratings around 6 out of 10.}
\label{fig:1b}
\end{minipage}%

\caption{Participants rated all panels as similar in relevance, regardless of number of relevant features included in the panel }\label{fig:figure1}
\end{figure*}

To design our experiment, we had to consider multiple trade-offs, which we discuss here.

\textbf{All panels were human-curated}:
Human-curated and algorithmically generated recommendations are likely to be qualitatively different. Our decision to make the panels identical, but changing only the purported source, allows us to isolate the impact of source alone, but does not capture these potential visual and semantic differences.

\textbf{Fixed Panels and Relevance}:
Four panels were computed in advance, based on a fixed set of attributes collected from our pilot study. Features such as visual encoding, information density, and attribute pairing were uniform across these panels, as we chose to vary only attribute relevance across these panels. While we considered tailoring panels based on participants' rankings, as well as varying the visual encodings, the strength of this design is that it keeps the number of experiment conditions at a manageable level. 
However, we acknowledge that attribute relevance is only one dimension of visualization quality, and that visual encodings and interaction effects of attribute pairings were not accounted for. We chose a simple relevance metric as a starting point rather than a more complex reactive measure or stimuli generation procedure. 

\textbf{Single Trial Design}:
In our experiment, a single trial consists of participants evaluating one pair of panels. While one recorded preference per participant limits statistical power, it also avoids bias from multiple explorations of the same dataset. It also allows the experiment to be easily crowdsourced to a platform like Prolific.

\textbf{Wide Range in Participant Expertise}:
By not constraining participation by experience, our user population may not accurately reflect the population that commonly use visualization recommendation features. Though another study with a different population would help generalize the results, this design allows us to directly observe how data analysis experience influences user recommendation preferences.

\textbf{Using Color to Distinguish Algorithm and Human Panels}:
Though color was randomly assigned as a way to distinguish between recommendation source, a small number of participants used it as a basis for their decision. In consideration of this subgroup, other means of distinguishing between algorithm versus human could have been used. 
\section{Analysis}
\label{sec:analysis}

In this section, we investigate our research questions via a mixture of quantitative and qualitative methods, focusing our analysis on the link between participants' prior preferences regarding recommendation sources, the participants' measured relevance of the displayed visualization panels, and their self-reported rationales for choosing one panel of visualization recommendations over another.

\subsection{Preliminaries}
In this section, we define specific concepts and calculations used throughout our analysis, and summarize our analysis methods. We compute panel relevance based on a fixed set of attributes collected in a pilot study, as well as by using participants' top five selected attributes (see \autoref{sec:experiment-design:relevance} and \autoref{sec:experiment-design:limitations}). The specific relevance measures we analyze are as follows:

\begin{description}[nosep]
  \itemsep0em
 \item[\emph{A Priori} or Participant Ranking] The ranking of the five most frequently selected attributes from our pilot study or by a specific participant respectively. 
\end{description}

\begin{description}[nosep]
 \item[Panel Rating] The rating that a participant assigned to a specific panel, using a ten point Likert scale. 
 \item[Attribute Overlap] Given two sets of attributes, we calculate the cardinality of the intersection of the two sets.

\end{description}
Note that each recommendation panel (and individual visualization within this panel) represents a unique subset of attributes, which may or may not overlap with a user's preferred attributes.

\subsubsection{Quantitative Analysis Methods Overview}
\label{sec:analysis:exploratory-analysis-overview}

Our quantitative measures aim to uniformly assess the decisions made by participants, such as the frequency of \highlight{when} each source was selected or the distribution of ratings assigned to each recommendation panel.
Given the context of our study design and tentative hypotheses, we also opted for a more exploratory rather than confirmatory design for analyzing our study data. Rather than relying on inferential statistical analysis, we used our quantitative analysis to identify patterns in participants' visualization preferences and analysis behaviors, and report on general effect sizes and confidence intervals to provide additional context for our findings.


\subsubsection{Qualitative Analysis Methods Overview}
\label{sec:analysis:qualitative-analysis-overview}
In our experiment, we explicitly asked participants to explain their choice in recommendation panel (see \autoref{sec:experiment-design:comparing}). These responses form the basis for our qualitative analysis. We note that two responses were excluded from the analysis. One response was not written in English, preventing an accurate evaluation, and the other was empty. We qualitatively coded 112 responses for this analysis.

We performed open coding~\cite{lazar2017research} on participants' free text responses explaining the reasons for their panel selection; the codes are provided in \autoref{tab:codes}.  To develop a consistent coding scheme, two researchers coded the first 40 responses individually, then discussed to resolve discrepancies.
One researcher coded the remaining responses, which the other reviewed. We used a multi-phase coding process to refine and merge similar codes. Then, the resulting 24 codes were organized into six high-level themes describing the reasoning behind participants' decision-making processes.

\subsection{Verifying Recommendation Relevance}
\label{sec:analysis:relevance}
Before evaluating the effects of recommendation source preference on decision-making, we assess the alignment between participants' perceptions of panel relevance and our calculated relevance scores. This analysis serves to not only anchor our measures, but also to explore our first hypothesis -- that people will rate recommendation sources according to the degree of attribute overlap. Our results are as follows:





\textbf{A priori rankings have high attribute overlap with participant rankings}:
We first analyze attribute overlap between a priori and participant rankings, shown in \autoref{fig:figure1}, where the range of attribute overlap is from 0 (i.e. no overlap) to 5 (i.e. both sets are identical).
We find similarities between a priori and participant rankings. 70\% of participants selected at least three of the same attributes used in the a priori rankings. This indicates that our a priori relevance seems to be aligned with participant relevance, and thus was a reasonable starting point for generating high- and low-relevance panels.

\begin{figure}
\centering
\begin{minipage}[t]{0.45\textwidth}
  \centering
\includegraphics[width=\textwidth]{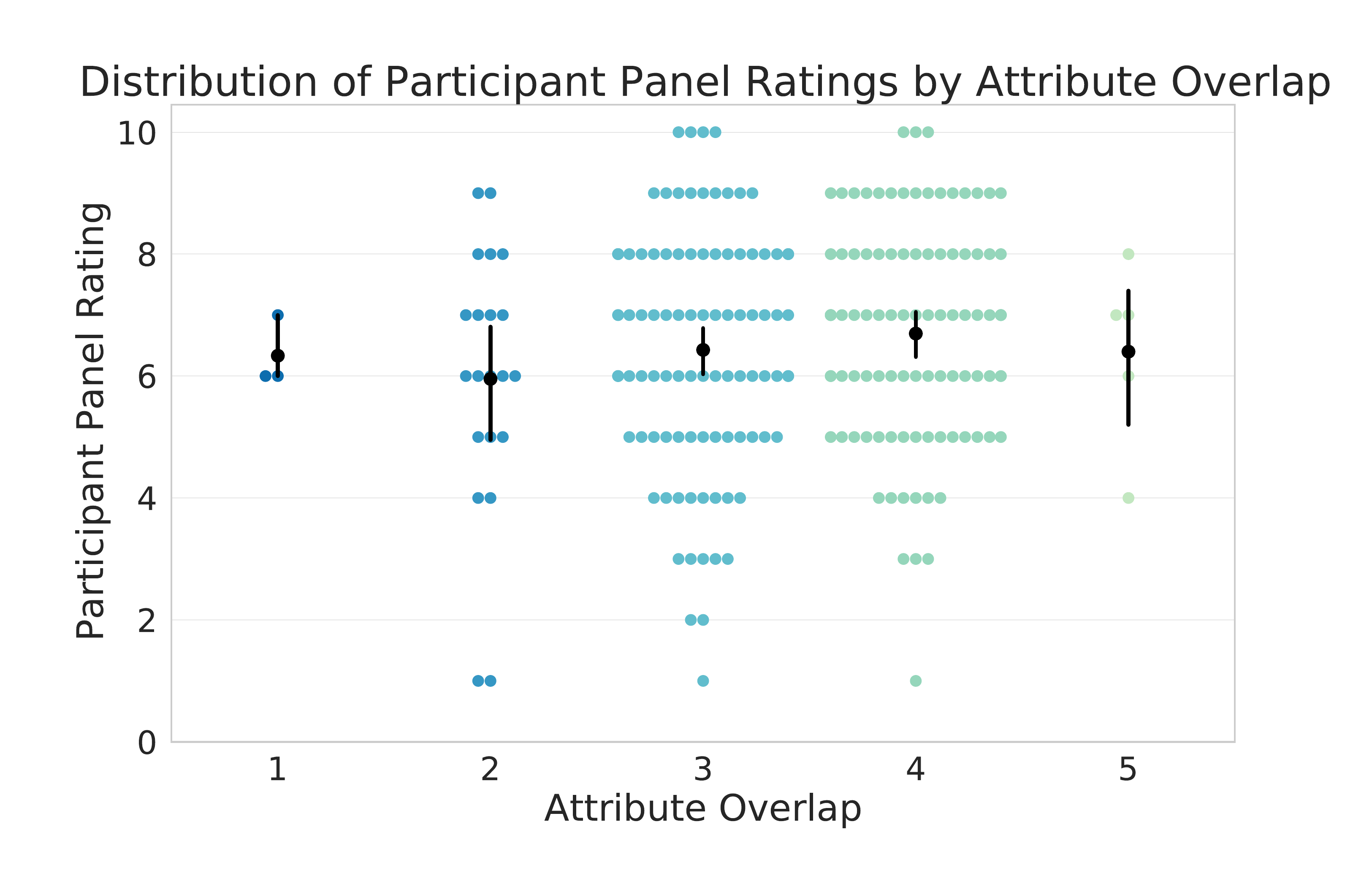}
\subcaption[first caption.]{A distribution of panel rating by attribute overlap: the number of data features selected as relevant by the participant that appear in a particular panel of recommendations. Bars represent 95\% confidence interval with 1000 bootstrapped iterations.}
\Description{Swarm plot of panel ratings, separated by attribute overlap. Despite different levels of attribute overlap, participants generally rated panels similarly.}
\end{minipage}
\begin{minipage}[t]{0.45\textwidth}
  \centering
\includegraphics[width=\textwidth]{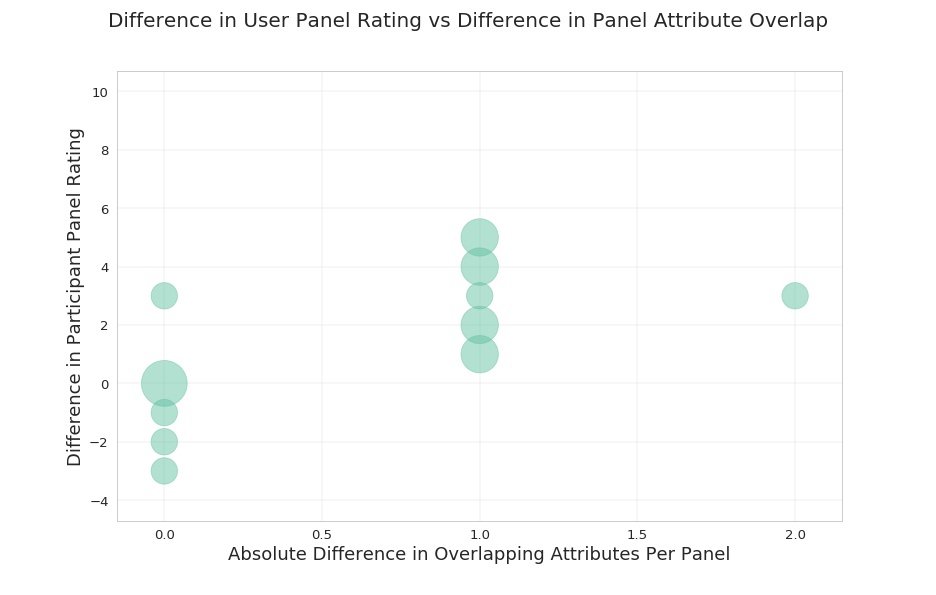}
\setlength{\belowcaptionskip}{-8pt}

\subcaption[second caption.]{
As comparatively more data fields selected by the participant as relevant appear in a panel of recommendations, the subjective rating of the chosen panel likewise increases. The y-axis is the comparative difference in rating between chosen and unchosen panels. The x-axis is the difference in attribute overlap between the chosen and unchosen panels, and the
radius is the number of responses.
}\label{fig:figures_2}
\Description{Bubble plot with the absolute difference in panel ratings vs absolute differences in attribute overlap. Bubble size encodes the number of participants with the corresponding score difference. There is an upward trend indicating that participants are giving increasingly different ratings when the panels exhibit increasingly different levels of attribute overlap.}
\end{minipage}
\setlength{\belowcaptionskip}{-8pt}

\caption{Relative difference appears to be a better metric than absolute difference.}\label{figures_1.png}
\end{figure}

\textbf{Participants seem to emphasize relative rather than absolute panel ratings}:

Participants rated the presented panels similarly. However, since participants were asked to compare \emph{pairs} of panels rather than panels in isolation, it is possible that participants are focusing on \emph{relative} differences when assigning ratings to panels. 
To evaluate this scenario, we first compare each panel to the corresponding ranking assigned by each participant. One panel will overlap more or equally with the participants' rankings than the other. Subtracting the lower overlap value from the higher produces a new value between zero and five. We then apply the corresponding calculation for panel ratings: subtracting the rating of the low-overlap panel from the rating of the high-overlap panel.
\autoref{fig:figures_2} illustrates the results, where the x-axis represents the difference in attribute overlap for a pair of panels, and the y-axis represents the corresponding difference in panel ratings. The figure illustrates a trend where as the difference in attribute overlap increases, so does the difference in panel rating. This suggests that participants may view relevance as a function of attribute overlap, and so rating alone may not be sufficient to describe the perceived utility of a panel. Thus, the panel ratings may be useful in terms of measuring \emph{relative} preference between a given pair of recommendation panels. 

\begin{table*}
\resizebox{\textwidth}{!}{%
\begin{tabular}{l|l|l|l|l}
\hline
\textbf{Themes} & \multicolumn{4}{l}{\textbf{Codes}} \\ \hline
Data-driven Decisions (60) & \multicolumn{4}{l}{Information quality (11), information usability (25), relevance (5), visualizations of interest (23), attributes of interest (27)} \\ 
Trust in Source (22) & \multicolumn{4}{l}{\begin{tabular}[c]{@{}l@{}}Trust in ability of humans or algorithm (19), human touch (3), trust in dataset (1), desire for insight on recommender's process (1)\end{tabular}} \\ 
Reliability of Source (21) & \multicolumn{4}{l}{Reliability (11), accuracy (10), errors (4)} \\ 
Participant Comprehension (17) & \multicolumn{4}{l}{Comprehension (16), ease of analysis (2)} \\ 
Personal Experiences (16) & \multicolumn{4}{l}{\begin{tabular}[c]{@{}l@{}}Personal preference (6), personal background (3), preference for a data representation (3), indifference (2)\end{tabular}} \\ 
Visual Aesthetics (7) & \multicolumn{4}{l}{Visual aesthetics (7), color preference (1)} \\ \hline
\end{tabular}
}
\caption{The codes derived from our qualitative analysis, organized by high-level themes. }
\Description{A 2 column table, with the first column containing all of the themes from our qualitative analysis and the frequency of occurrence. The second column contains the codes associated with that theme, as well as their frequency of occurrence. The 6 themes are Data-driven decisions, Trust in Source, Reliability of Source, Participant Comprehension, Personal Experiences, and Visual Aesthetics.}
\label{tab:codes}
\end{table*}

\subsection{Assessing Bias in Participant Preferences and Panel Selections}
\label{sec:analysis:bias}



In this section, we explore our first research question:
\emph{How do existing preferences for human-curated versus algorithmically-generated recommendations affect the evaluation of recommendation quality or utility?} This  will also help us explore our second hypothesis -- \emph{that an a priori preference towards a particular recommendation source predisposes participants to select a particular panel, regardless of recommendation relevance.} 

\begin{figure}[htb]
    \centering
    \includegraphics[width=\columnwidth]{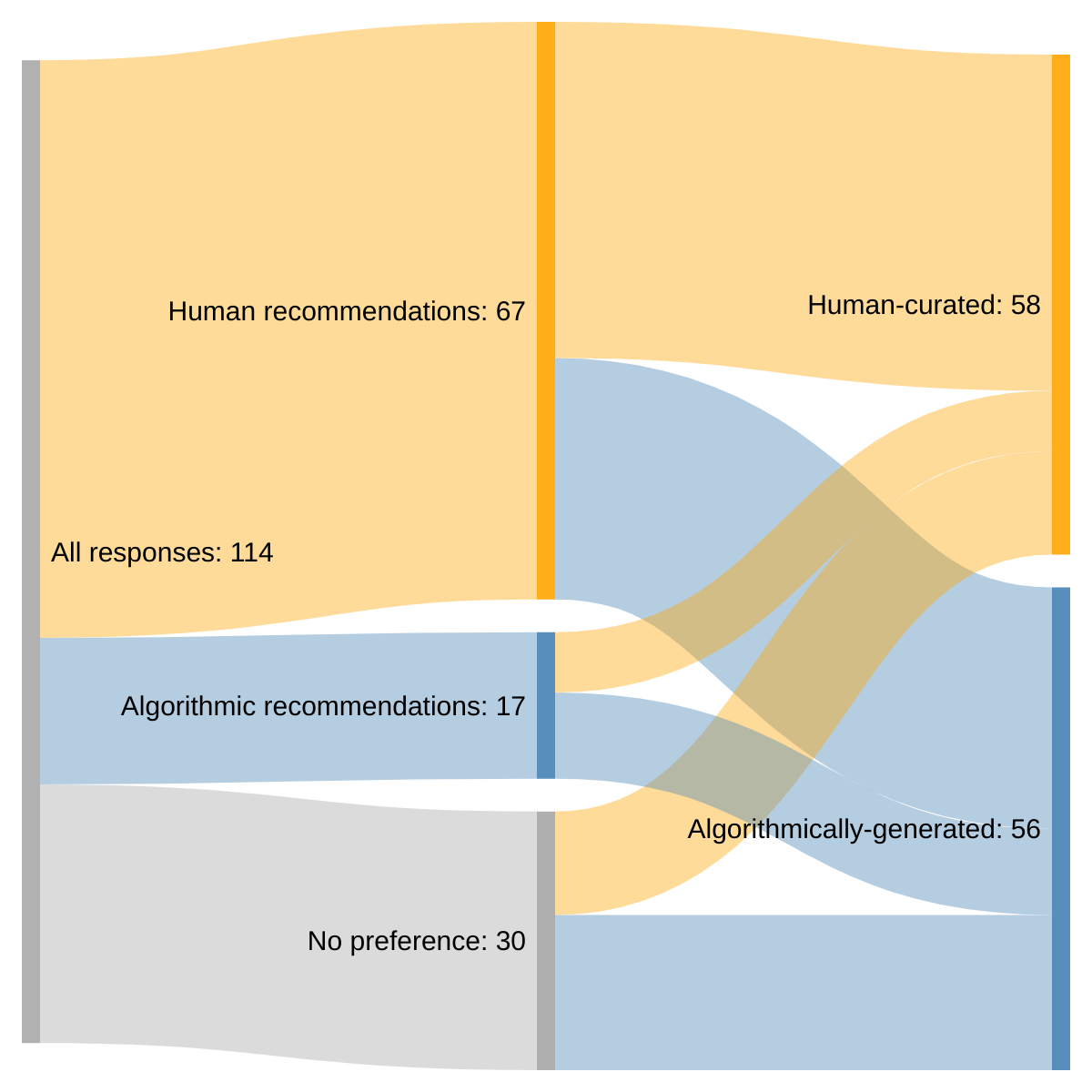}
    \setlength{\belowcaptionskip}{-8pt}
    \caption{Sankey diagram of prior preferences of recommendation sources and posterior selection of panels. Most people expressed a prior preference for human sources, but the resulting posteriors are split nearly 50/50}
    \label{fig:sankey}
    \Description{Sankey diagram of prior preferences of recommendation sources and posterior selection of panels. Most people expressed a prior preference for human sources, but the resulting posteriors are split nearly 50/50}
\end{figure}

\textbf{Though participants initially prefer human recommendations, they are neutral to source in their panel selections}: First, we delved into understanding the existing preferences and subsequent decisions made by our participants. 60\% indicate a prior preference for human recommendations, 15\% prefer algorithmic recommendations, and 25\% have no preference. Though this suggests an existing preference for human recommendation sources, \highlight{participants'} subsequent recommendation source choices are evenly split between humans and algorithms (see \autoref{fig:sankey}).

\textbf{Prior preferences are not highly predictive of visualization recommendation choices:} For participants who initially \highlight{preferred} human recommendations, 58\% \highlight{selected} the ``Human-Curated'' panel. For participants who initially \highlight{preferred} algorithmic recommendations, 59\% \highlight{selected} the ``Algorithmically-Generated'' panel. 62\% of participants with no initial preference selected the ``Algorithmically-Generated'' panel. Given that we controlled for relevance across all conditions, these findings suggest that participants \highlight{did} not primarily select panels based on their a priori preference for human or algorithmic recommendations. Nevertheless, there was a subset of participants that did not follow this trend.  

\textbf{A fraction of adherents chose their panel despite estimated lower relevance}: We find that 14\% of all participants stick with their prior preferences, even when the calculated relevance of the chosen panel is lower than that of the unchosen panel (average relevance of chosen of 8, unchosen is 11). Though far from a majority, this result may suggest that some participants may still make allowances for their preferred recommendation source, and thus may be susceptible to recommendation errors.

It is possible that our relevance metric did not align with how these specific participants' models of relevance. Our metric was calculated using participants' attribute rankings only, and does not account for interaction effects between attributes or other potential indicators of utility. As a result, we qualitatively explore alternative measures of relevance from participants in the next section.

In summary, though our participants initially appeared to have preferences for human recommendation sources, they were ultimately neutral to visualization recommendation sources. However, there is a minority of people that appear to make allowances for their prior preferences, which may drive their choice of recommendation source. 
%

\subsection{Understanding Participants' Decision Making Models}

After observing that most participants claim to prefer human recommendations yet in practice are insensitive towards recommendation source, we explore our second research question: \emph{If clear preferences are observed, what reasons do users give for preferring a certain recommendation source?} We qualitatively analyze participant responses to understand their reasoning for choosing a certain panel.

\textbf{Many participants reported selecting panels in an analysis-driven way}: The most common theme (60 out of 112 responses) was analysis-driven decisions, meaning that participants evaluated panels based on their own analysis interests and the corresponding applicability of the information encoded within the visualizations. Within that umbrella category, participants fell into different clusters of reasoning. Some participants seemed to regard each panel as a whole and chose a panel based on the combined utility of the component visualizations. As an example, one participant noted: 
\begin{quote}
    \emph{Although the selection doesnt reflect all of the attributes picked earlier on, i think the data in this set matches more so what i want to focus on in the pitch. (P91)}
\end{quote}
\vspace{-1mm}
These participants seem to suggest that they prioritized the cumulative information gained when choosing one panel versus the other.
On the other hand, there were participants who noted specific visualizations or attributes that swayed their decision to choose a particular panel, and thus recommendation source. In particular, 19 of these 60 participants mention specific visualizations within their justification. For example, one participant wrote: 
\vspace{-1mm}
\begin{quote}
    \emph{[the] last bit of human curated (domestic gross v release year) was really convincing chart (P33)}
\end{quote}
\vspace{-1mm}

Lastly, many participants expressed interest in specific attributes or visualizations, situated with respect to the context of their movie pitches. For example, another participant stated:
\vspace{-1mm}
\begin{quote}
\emph{[the] algorithmically-generated grid has some very useful graphs, such as profit/mpaa rating and budget/genre...the company that the movie is being pitched to would be really interested in these particular sets of data. (P54)}
\end{quote}
\vspace{-1mm}
Overall, even within the data-driven decisions theme, there was a significant diversity in rationales, be it the combined utility of visualizations, a singular visualization, or focusing on potentially interesting attributes for the target audience.

\textbf{Trust in the source of recommendations was the second most common reason for choosing a particular panel}. In examining adherents in \autoref{sec:analysis:bias}, we became interested in understanding the potential biases people harbor towards a particular recommendation source. We observed the theme ``Trust in Source'' in 22 of 112 responses. 

Some participants believed that algorithms are more efficient and less likely to make errors than humans (6 of 22 responses). As P75 explained, 
\vspace{-1mm}
\begin{quote}
   \emph{``Although they [the panels] both seemed equally helpful to me, I chose the algorithmical one because human curated may have some mistakes in it but algorithmical one most probably does not.''}
\end{quote}
\vspace{-1mm}

This reasoning came from a general impression that algorithms are more accurate, or from the participant's personal background, as one participant notes: \emph{...I inhere[n]tly tend to trust Algorithms more since I am a computer s[c]ience student. (P97)}

Conversely, some participants indicated a preference for humans due to the presence of a ``human touch'' in the recommendation panels (3 of 22 responses). For example, one participant
references this notion of a uniquely human quality in the human-curated visualizations by explaining that
\vspace{-1mm}
\begin{quote}
   \emph{``Both [panels] had good information but I think the human touch is required to get the best feel for the data'' (P65).} 
\end{quote}
\vspace{-1mm}
Along a similar vein, one participant made their choice based upon the notion that \emph{because} a human recommended these visualizations, it had intrinsic value:
\begin{quote}
\emph{``I prefer things made by actual people and not algorithms, although the later could be useful too'' (P94)}
\end{quote}
\vspace{-1mm}
When both recommendation options are perceived as equally relevant, factors outside of the data relevance become increasingly important. Some participants will incorporate prior preferences for algorithmic or human recommendations into their decision making process. In the case of participant P65, the ``human touch'' becomes the deciding factor in selecting the ``Human-Curated'' panel. For participant P75, it was the added legitimacy and accuracy that an algorithm lends to recommendations. 

\subsection{Exploring How Analysis Experience And Other Factors Influence Participant Preferences}
Given the wide range of factors that could affect a participant's evaluation strategies, we seek to better understand how data analysis experience may influence how these factors are prioritized by participants.
In this section, we explore our third research question: \emph{What effect (if any) does statistical or data analysis experience have on user preferences?}

\textbf{Analysis experience does not seem to impact prior preference or panel selections}:
We recorded four measures of data analysis experience: statistics, visualization, general data analysis, and data analysis tool (Excel, R, Matlab, etc.) experience. For all measures, we failed to find a notable difference across experience levels. For sake of space, we only present a single salient example.  
We \highlight{found} that participants selected the ``Algorithmically-Generated'' panel at similar rates regardless of frequency of performing data analysis tasks, used as an indicator for general data analysis experience level: 55\% (almost never), 53\%(less than monthly), 50\% (less than weekly), 33\%(weekly), and 67\% (daily). Thus, data analysis experience does not appear to be predictive of panel selection, answering our third research question. 

\textbf{Visualization comprehension may affect participants' ability to evaluate recommendations}: Despite this, a notable number of responses were categorized under ``Participant Comprehension'' (17 of 112 responses).
In particular, we find that some participants had trouble interpreting the visualizations in some of the panels. For example, one participant mentions that \emph{``the ones with the little dots are very confusing'' (P61).} This issue touches on one of the limitations of our experiment design (see \autoref{sec:experiment-design:limitations}), and poses a challenge for recommendations in general: not everyone values the relevance of a recommendation in the same way, and relevance can mean different things for users of differing levels of experience. For example, novice users may prioritize comprehension and ease of understanding, whereas more experienced users may prioritize insight density.


\textbf{Aesthetics can be a deciding factor in choosing between visualization recommendation panels}: The aesthetics of visualizations also seems to be a relevant factor in judging recommendations. Seven of 112 responses contained the  ``Visual Aesthetics'' theme, which refers to the artistic appearance of the visualizations themselves. Two participants cited the ``cleanness'' and ``simplicity'' of certain panels as the reasons for their selections. One participant in particular selected their chosen panel because the \emph{``blue color is visually more impactful'' (P55).} These sorts of design choices are often overlooked in making recommendations, to the detriment of users who value them.



Based on these findings, we summarize how participants seem to choose between panels: participants often approach the evaluation of visualization recommendations in an analysis-driven way, but an inherent trust in recommendation source may play a role as well. For some, comprehension and visual aesthetics were also notable determining factors in selecting specific recommendation panels.

\section{Discussion}

From our quantitative and qualitative observations, we synthesize the following preliminary findings: 

\textbf{People seemed to choose panels by focusing on the information presented in the visualizations.} In examining the rationale behind panel choices we found that, for a majority of participants, their decision-making process often focuses on the relevance of the data itself. We found some evidence that that \emph{people may rate recommendation panels based on the level of attribute overlap}. For example, participants may emphasize the cumulative information gained from a panel. We refer to these participants as \textit{all-rounders}, as they seem to obtain value from the quantity of information in a panel. However, some participants rely on a particular visualization as a deciding factor, indicating that prior preferences (and therefore biases) may still play a role in subsequent decision-making. We refer to this cohort as \textit{seekers}, as they seem to be looking for a specific chart or set of charts to investigate specific hypotheses.

Our results point to the existence of differing patterns of information foraging among viewers, suggesting that there may not be a one-size-fits-all solution to the problem of recommending useful visualizations. To better support these different user groups, we suggest tailoring recommendations towards different patterns of analyses. For instance, recommendations for all-rounders could focus on presenting dashboards that present overviews of many key metrics, whereas recommendations for seekers could focus on specific attributes of interest or particularly informative visualizations (in the information-theory sense~\cite{chen2010information}). These suggestions might require the collection of priors or other data-driven inference about data, rather than just the surfacing of arbitrary ``data facts''~\cite{srinivasan2018augmenting} such as outliers or strong correlations.



\textbf{Existing preferences are not predictive of subsequent recommendation choice, though perceptions of source can contribute.} While participants initially
preferred human recommendations, their subsequent panel choices revealed an insensitivity to recommendation source. Similarly, Peck et al. found that over half of participants did not change their initial rankings of the usefulness of visualizations after the sources (e.g., government agencies, universities) of each visualization were revealed~\cite{peck2019data}. That said, around a fifth of participants in our study indicated that the source of recommendations was important to them, and some perceived differences in the reliability and accuracy of human and algorithmic recommendations.  

Designers should consider how to present information in an unbiased manner for these individuals with clear preferences for recommendation source. For example, visualization recommendation systems might provide more detailed information on how the recommendations were curated to help users develop more trust in the system as a whole ~\cite{YangVisualExplanations}. However, more research is needed to determine an appropriate threshold for transparency: providing too much or too little transparency can erode trust in algorithmic recommendations~\cite{Kizilcec2016}.





\textbf{Data analysis experience does not have a substantial impact on preferences or panel selections, although visualization comprehension and aesthetics may play a role}. Challenges in understanding certain visualizations (e.g., scatterplots) may have hindered some participants' ability to effectively evaluate recommendations. Although a recommendation panel may have useful information, it is only relevant if the user can meaningfully interpret its data. Future work may solicit more information about participants' personal backgrounds to gain deeper insight into factors that influence people's perceptions of relevance beyond data analysis experience. While analysis experience did not seem to impact participants' panel selections, aesthetics shaped a minority of participants' decisions. As the audience of visualization tools broadens over time, designers should consider the aesthetics of visualizations as a means of supporting users' understanding of the data and encouraging them to interact with the system \cite{moere2011role}.

Overall, our results are positive news for the designers of visualization recommendation systems. Neither of the failure cases we mention in this paper, either the uncritical acceptance of recommendations from algorithmic sources, nor the knee-jerk rejection of the same, appear to occur with high frequency. However, we note that this pattern of apparent rationality is not universal: a sizable portion of our adherents stuck with their preferred choice of recommendation source even given measurable mismatches in expected utility. This behavior suggests that we either lack appropriate measures to assess the utility of visualization recommendations, or that users rely on additional factors to evaluate perceived differences between recommenders. As a consequence, future work may investigate the qualitative and quantitative differences between human and algorithmic data visualizations to identify specific factors influencing trust. The presence of these adherents, combined with a substantial initial preference for human recommendations, also suggests that there is value (and some risk) in providing human recommendations, which may be perceived as having intrinsic authority and relevance over algorithmic choices. 

\section{Conclusion}
As visualization systems increasingly rely on automated or semi-automated methods, and as the population of people who encounter visualization tools becomes larger and less specialized, the attitudes and beliefs of our users towards algorithmic recommendations will become increasingly important. The results of our experiment show that, for the most part, people seem to assess visualization recommendations in terms of relevance, rather than source; this seemed to hold across groups of varying statistical experience. However, there was a minority of people who did not seem to act in a data driven way. For instance, visual comprehension and aesthetics choices seemed to drive the decision-making of some participants, who should also be considered when designing large-scale systems. In that way, people are (with some exceptions) generally capable of adjusting their beliefs about the source of recommendations to fit their analytical needs, and making informed decisions about what recommendations to trust.
\section{Acknowledgements}
We thank all the reviewers, study participants, and members of both the Human-Computer Interaction Lab and the Battle Data Lab for their valuable feedback.

\bibliographystyle{ACM-Reference-Format}

\bibliography{references.bib}


\begin{thebibliography}{39}


\ifx \showCODEN    \undefined \def \showCODEN     #1{\unskip}     \fi
\ifx \showDOI      \undefined \def \showDOI       #1{#1}\fi
\ifx \showISBNx    \undefined \def \showISBNx     #1{\unskip}     \fi
\ifx \showISBNxiii \undefined \def \showISBNxiii  #1{\unskip}     \fi
\ifx \showISSN     \undefined \def \showISSN      #1{\unskip}     \fi
\ifx \showLCCN     \undefined \def \showLCCN      #1{\unskip}     \fi
\ifx \shownote     \undefined \def \shownote      #1{#1}          \fi
\ifx \showarticletitle \undefined \def \showarticletitle #1{#1}   \fi
\ifx \showURL      \undefined \def \showURL       {\relax}        \fi
\providecommand\bibfield[2]{#2}
\providecommand\bibinfo[2]{#2}
\providecommand\natexlab[1]{#1}
\providecommand\showeprint[2][]{arXiv:#2}

\bibitem[\protect\citeauthoryear{Bresciani and Eppler}{Bresciani and
  Eppler}{2015}]%
        {bresciani2015pitfalls}
\bibfield{author}{\bibinfo{person}{Sabrina Bresciani} {and}
  \bibinfo{person}{Martin~J Eppler}.} \bibinfo{year}{2015}\natexlab{}.
\newblock \showarticletitle{The pitfalls of visual representations: A review
  and classification of common errors made while designing and interpreting
  visualizations}.
\newblock \bibinfo{journal}{\emph{Sage Open}} \bibinfo{volume}{5},
  \bibinfo{number}{4} (\bibinfo{year}{2015}),
  \bibinfo{pages}{2158244015611451}.
\newblock
\urldef\tempurl%
\url{https://doi.org/10.1177/2158244015611451}
\showDOI{\tempurl}


\bibitem[\protect\citeauthoryear{Castelo, Bos, and Lehmann}{Castelo
  et~al\mbox{.}}{2019}]%
        {castelo2019let}
\bibfield{author}{\bibinfo{person}{Noah Castelo}, \bibinfo{person}{Maarten~W
  Bos}, {and} \bibinfo{person}{Donald Lehmann}.}
  \bibinfo{year}{2019}\natexlab{}.
\newblock \showarticletitle{Let the Machine Decide: When Consumers Trust or
  Distrust Algorithms}.
\newblock \bibinfo{journal}{\emph{NIM Marketing Intelligence Review}}
  \bibinfo{volume}{11}, \bibinfo{number}{2} (\bibinfo{year}{2019}),
  \bibinfo{pages}{24--29}.
\newblock
\urldef\tempurl%
\url{https://doi.org/10.2478/nimmir-2019-0012}
\showDOI{\tempurl}


\bibitem[\protect\citeauthoryear{Chen and Jaenicke}{Chen and Jaenicke}{2010}]%
        {chen2010information}
\bibfield{author}{\bibinfo{person}{Min Chen} {and} \bibinfo{person}{Heike
  Jaenicke}.} \bibinfo{year}{2010}\natexlab{}.
\newblock \showarticletitle{An information-theoretic framework for
  visualization}.
\newblock \bibinfo{journal}{\emph{IEEE Transactions on Visualization and
  Computer Graphics}} \bibinfo{volume}{16}, \bibinfo{number}{6}
  (\bibinfo{year}{2010}), \bibinfo{pages}{1206--1215}.
\newblock
\urldef\tempurl%
\url{https://doi.org/10.1109/TVCG.2010.131}
\showDOI{\tempurl}


\bibitem[\protect\citeauthoryear{Cheng, Ming, and Qu}{Cheng
  et~al\mbox{.}}{2020}]%
        {cheng2020dece}
\bibfield{author}{\bibinfo{person}{Furui Cheng}, \bibinfo{person}{Yao Ming},
  {and} \bibinfo{person}{Huamin Qu}.} \bibinfo{year}{2020}\natexlab{}.
\newblock \bibinfo{title}{DECE: Decision Explorer with Counterfactual
  Explanations for Machine Learning Models}.
\newblock
\newblock
\showeprint[arxiv]{2008.08353}~[cs.LG]


\bibitem[\protect\citeauthoryear{Correll}{Correll}{2019}]%
        {correllEthical}
\bibfield{author}{\bibinfo{person}{Michael Correll}.}
  \bibinfo{year}{2019}\natexlab{}.
\newblock \showarticletitle{Ethical Dimensions of Visualization Research}. In
  \bibinfo{booktitle}{\emph{Proceedings of the 2019 CHI Conference on Human
  Factors in Computing Systems}}. \bibinfo{publisher}{Association for Computing
  Machinery}, \bibinfo{address}{New York, NY, USA}, \bibinfo{pages}{1–13}.
\newblock
\showISBNx{9781450359702}
\urldef\tempurl%
\url{https://doi.org/10.1145/3290605.3300418}
\showDOI{\tempurl}


\bibitem[\protect\citeauthoryear{Dibia and Demiralp}{Dibia and
  Demiralp}{2019}]%
        {dibia2018data2vis}
\bibfield{author}{\bibinfo{person}{Victor Dibia} {and}
  \bibinfo{person}{{\c{C}}a{\u{g}}atay Demiralp}.}
  \bibinfo{year}{2019}\natexlab{}.
\newblock \showarticletitle{Data2Vis: Automatic generation of data
  visualizations using sequence to sequence recurrent neural networks}.
\newblock \bibinfo{journal}{\emph{IEEE Computer Graphics and Applications}}
  \bibinfo{volume}{39}, \bibinfo{number}{5} (\bibinfo{year}{2019}),
  \bibinfo{pages}{33--46}.
\newblock
\urldef\tempurl%
\url{https://doi.org/10.1109/MCG.2019.2924636}
\showDOI{\tempurl}


\bibitem[\protect\citeauthoryear{Edwards, Edwards, Spence, and Shelton}{Edwards
  et~al\mbox{.}}{2014}]%
        {edwards2014bot}
\bibfield{author}{\bibinfo{person}{Chad Edwards}, \bibinfo{person}{Autumn
  Edwards}, \bibinfo{person}{Patric~R Spence}, {and}
  \bibinfo{person}{Ashleigh~K Shelton}.} \bibinfo{year}{2014}\natexlab{}.
\newblock \showarticletitle{Is that a bot running the social media feed?
  Testing the differences in perceptions of communication quality for a human
  agent and a bot agent on Twitter}.
\newblock \bibinfo{journal}{\emph{Computers in Human Behavior}}
  \bibinfo{volume}{33} (\bibinfo{year}{2014}), \bibinfo{pages}{372--376}.
\newblock
\urldef\tempurl%
\url{https://doi.org/10.1016/j.chb.2013.08.013}
\showDOI{\tempurl}


\bibitem[\protect\citeauthoryear{Golfarelli and Rizzi}{Golfarelli and
  Rizzi}{2020}]%
        {golfarelli2019model}
\bibfield{author}{\bibinfo{person}{Matteo Golfarelli} {and}
  \bibinfo{person}{Stefano Rizzi}.} \bibinfo{year}{2020}\natexlab{}.
\newblock \showarticletitle{A model-driven approach to automate data
  visualization in big data analytics}.
\newblock \bibinfo{journal}{\emph{Information Visualization}}
  \bibinfo{volume}{19}, \bibinfo{number}{1} (\bibinfo{year}{2020}),
  \bibinfo{pages}{24--47}.
\newblock
\urldef\tempurl%
\url{https://doi.org/10.1177/1473871619858933}
\showDOI{\tempurl}


\bibitem[\protect\citeauthoryear{Graefe, Haim, Haarmann, and Brosius}{Graefe
  et~al\mbox{.}}{2016}]%
        {2016newscredibility}
\bibfield{author}{\bibinfo{person}{Andreas Graefe}, \bibinfo{person}{Mario
  Haim}, \bibinfo{person}{Bastian Haarmann}, {and} \bibinfo{person}{Hans-Bernd
  Brosius}.} \bibinfo{year}{2016}\natexlab{}.
\newblock \showarticletitle{Perception of Automated Computer-Generated News:
  Credibility, Expertise, and Readability}.
\newblock \bibinfo{journal}{\emph{Journalism}}  \bibinfo{volume}{19}
  (\bibinfo{year}{2016}).
\newblock
\urldef\tempurl%
\url{https://doi.org/10.1177/1464884916641269}
\showDOI{\tempurl}


\bibitem[\protect\citeauthoryear{Heer}{Heer}{2019}]%
        {heer2019agency}
\bibfield{author}{\bibinfo{person}{Jeffrey Heer}.}
  \bibinfo{year}{2019}\natexlab{}.
\newblock \showarticletitle{Agency plus automation: Designing artificial
  intelligence into interactive systems}.
\newblock \bibinfo{journal}{\emph{Proceedings of the National Academy of
  Sciences}} \bibinfo{volume}{116}, \bibinfo{number}{6} (\bibinfo{year}{2019}),
  \bibinfo{pages}{1844--1850}.
\newblock
\urldef\tempurl%
\url{https://doi.org/10.1073/pnas.1807184115}
\showDOI{\tempurl}


\bibitem[\protect\citeauthoryear{Jakesch, French, Ma, Hancock, and
  Naaman}{Jakesch et~al\mbox{.}}{2019}]%
        {jakesch2019ai}
\bibfield{author}{\bibinfo{person}{Maurice Jakesch}, \bibinfo{person}{Megan
  French}, \bibinfo{person}{Xiao Ma}, \bibinfo{person}{Jeffrey~T. Hancock},
  {and} \bibinfo{person}{Mor Naaman}.} \bibinfo{year}{2019}\natexlab{}.
\newblock \showarticletitle{AI-Mediated Communication: How the Perception That
  Profile Text Was Written by AI Affects Trustworthiness}. In
  \bibinfo{booktitle}{\emph{Proceedings of the 2019 CHI Conference on Human
  Factors in Computing Systems}}. \bibinfo{publisher}{Association for Computing
  Machinery}, \bibinfo{address}{New York, NY, USA}, \bibinfo{pages}{1–13}.
\newblock
\showISBNx{9781450359702}
\urldef\tempurl%
\url{https://doi.org/10.1145/3290605.3300469}
\showDOI{\tempurl}


\bibitem[\protect\citeauthoryear{Kaggle}{Kaggle}{2019a}]%
        {kaggle1}
\bibfield{author}{\bibinfo{person}{Kaggle}.} \bibinfo{year}{2019 (accessed
  December 5, 2019)}\natexlab{a}.
\newblock \bibinfo{title}{IMDB 5000 Movie Dataset}.
\newblock
\newblock
\urldef\tempurl%
\url{https://www.kaggle.com/carolzhangdc/imdb-5000-movie-dataset}
\showURL{%
\tempurl}


\bibitem[\protect\citeauthoryear{Kaggle}{Kaggle}{2019b}]%
        {kaggle2}
\bibfield{author}{\bibinfo{person}{Kaggle}.} \bibinfo{year}{2019 (accessed
  December 5, 2019)}\natexlab{b}.
\newblock \bibinfo{title}{The Movies Dataset}.
\newblock
\newblock
\urldef\tempurl%
\url{https://www.kaggle.com/rounakbanik/the-movies-dataset}
\showURL{%
\tempurl}


\bibitem[\protect\citeauthoryear{Kaggle}{Kaggle}{2019c}]%
        {kaggle3}
\bibfield{author}{\bibinfo{person}{Kaggle}.} \bibinfo{year}{2019 (accessed
  December 5, 2019)}\natexlab{c}.
\newblock \bibinfo{title}{The Story of Film}.
\newblock
\newblock
\urldef\tempurl%
\url{https://www.kaggle.com/rounakbanik/the-story-of-film}
\showURL{%
\tempurl}


\bibitem[\protect\citeauthoryear{Kizilcec}{Kizilcec}{2016}]%
        {Kizilcec2016}
\bibfield{author}{\bibinfo{person}{Ren\'{e}~F. Kizilcec}.}
  \bibinfo{year}{2016}\natexlab{}.
\newblock \showarticletitle{How Much Information? Effects of Transparency on
  Trust in an Algorithmic Interface}. In \bibinfo{booktitle}{\emph{Proceedings
  of the 2016 CHI Conference on Human Factors in Computing Systems}}.
  \bibinfo{publisher}{Association for Computing Machinery},
  \bibinfo{address}{New York, NY, USA}, \bibinfo{pages}{2390–2395}.
\newblock
\showISBNx{9781450333627}
\urldef\tempurl%
\url{https://doi.org/10.1145/2858036.2858402}
\showDOI{\tempurl}


\bibitem[\protect\citeauthoryear{Lazar, Feng, and Hochheiser}{Lazar
  et~al\mbox{.}}{2017}]%
        {lazar2017research}
\bibfield{author}{\bibinfo{person}{Jonathan Lazar},
  \bibinfo{person}{Jinjuan~Heidi Feng}, {and} \bibinfo{person}{Harry
  Hochheiser}.} \bibinfo{year}{2017}\natexlab{}.
\newblock \bibinfo{booktitle}{\emph{Research methods in human-computer
  interaction}}.
\newblock \bibinfo{publisher}{Morgan Kaufmann}, \bibinfo{address}{San
  Francisco}.
\newblock


\bibitem[\protect\citeauthoryear{Lee}{Lee}{2018}]%
        {lee2018understanding}
\bibfield{author}{\bibinfo{person}{Min~Kyung Lee}.}
  \bibinfo{year}{2018}\natexlab{}.
\newblock \showarticletitle{Understanding perception of algorithmic decisions:
  Fairness, trust, and emotion in response to algorithmic management}.
\newblock \bibinfo{journal}{\emph{Big Data \& Society}} \bibinfo{volume}{5},
  \bibinfo{number}{1} (\bibinfo{year}{2018}),
  \bibinfo{pages}{2053951718756684}.
\newblock
\urldef\tempurl%
\url{https://doi.org/10.1177/2053951718756684}
\showDOI{\tempurl}


\bibitem[\protect\citeauthoryear{M.~Logg, A.~Minson, and A.~Moore}{M.~Logg
  et~al\mbox{.}}{2019}]%
        {logg}
\bibfield{author}{\bibinfo{person}{Jennifer M.~Logg}, \bibinfo{person}{Julia
  A.~Minson}, {and} \bibinfo{person}{Don A.~Moore}.}
  \bibinfo{year}{2019}\natexlab{}.
\newblock \showarticletitle{Algorithm appreciation: People prefer algorithmic
  to human judgment}.
\newblock \bibinfo{journal}{\emph{Organizational Behavior and Human Decision
  Processes}}  \bibinfo{volume}{151} (\bibinfo{year}{2019}),
  \bibinfo{pages}{90--103}.
\newblock
\urldef\tempurl%
\url{https://doi.org/10.1016/j.obhdp.2018.12.005}
\showDOI{\tempurl}


\bibitem[\protect\citeauthoryear{Mackinlay, Hanrahan, and Stolte}{Mackinlay
  et~al\mbox{.}}{2007}]%
        {mackinlay2007show}
\bibfield{author}{\bibinfo{person}{Jock Mackinlay}, \bibinfo{person}{Pat
  Hanrahan}, {and} \bibinfo{person}{Chris Stolte}.}
  \bibinfo{year}{2007}\natexlab{}.
\newblock \showarticletitle{Show me: Automatic presentation for visual
  analysis}.
\newblock \bibinfo{journal}{\emph{IEEE Transactions on Visualization and
  Computer Graphics}} \bibinfo{volume}{13}, \bibinfo{number}{6}
  (\bibinfo{year}{2007}), \bibinfo{pages}{1137--1144}.
\newblock
\urldef\tempurl%
\url{https://doi.org/10.1109/TVCG.2007.70594}
\showDOI{\tempurl}


\bibitem[\protect\citeauthoryear{Merritt, Heimbaugh, LaChapell, and
  Lee}{Merritt et~al\mbox{.}}{2013}]%
        {merrittAmbiguous}
\bibfield{author}{\bibinfo{person}{Stephanie~M. Merritt},
  \bibinfo{person}{Heather Heimbaugh}, \bibinfo{person}{Jennifer LaChapell},
  {and} \bibinfo{person}{Deborah Lee}.} \bibinfo{year}{2013}\natexlab{}.
\newblock \showarticletitle{I Trust It, but I Don’t Know Why: Effects of
  Implicit Attitudes Toward Automation on Trust in an Automated System}.
\newblock \bibinfo{journal}{\emph{Human Factors}} \bibinfo{volume}{55},
  \bibinfo{number}{3} (\bibinfo{year}{2013}), \bibinfo{pages}{520--534}.
\newblock
\urldef\tempurl%
\url{https://doi.org/10.1177/0018720812465081}
\showDOI{\tempurl}


\bibitem[\protect\citeauthoryear{Microsoft}{Microsoft}{2019}]%
        {powerbi}
\bibfield{author}{\bibinfo{person}{Microsoft}.} \bibinfo{year}{2019 (accessed
  December 5, 2019)}\natexlab{}.
\newblock \bibinfo{title}{Generate data insights automatically with Power BI}.
\newblock
\newblock
\urldef\tempurl%
\url{https://docs.microsoft.com/en-us/power-bi/service-insights}
\showURL{%
\tempurl}


\bibitem[\protect\citeauthoryear{Moere and Purchase}{Moere and
  Purchase}{2011}]%
        {moere2011role}
\bibfield{author}{\bibinfo{person}{Andrew~Vande Moere} {and}
  \bibinfo{person}{Helen Purchase}.} \bibinfo{year}{2011}\natexlab{}.
\newblock \showarticletitle{On the role of design in information
  visualization}.
\newblock \bibinfo{journal}{\emph{Information Visualization}}
  \bibinfo{volume}{10}, \bibinfo{number}{4} (\bibinfo{year}{2011}),
  \bibinfo{pages}{356--371}.
\newblock
\urldef\tempurl%
\url{https://doi.org/10.1177/1473871611415996}
\showDOI{\tempurl}


\bibitem[\protect\citeauthoryear{Moritz, Wang, Nelson, Lin, Smith, Howe, and
  Heer}{Moritz et~al\mbox{.}}{2018}]%
        {moritz2018formalizing}
\bibfield{author}{\bibinfo{person}{Dominik Moritz}, \bibinfo{person}{Chenglong
  Wang}, \bibinfo{person}{Greg~L Nelson}, \bibinfo{person}{Halden Lin},
  \bibinfo{person}{Adam~M Smith}, \bibinfo{person}{Bill Howe}, {and}
  \bibinfo{person}{Jeffrey Heer}.} \bibinfo{year}{2018}\natexlab{}.
\newblock \showarticletitle{Formalizing visualization design knowledge as
  constraints: Actionable and extensible models in Draco}.
\newblock \bibinfo{journal}{\emph{IEEE Transactions on Visualization and
  Computer Graphics}} \bibinfo{volume}{25}, \bibinfo{number}{1}
  (\bibinfo{year}{2018}), \bibinfo{pages}{438--448}.
\newblock
\urldef\tempurl%
\url{https://doi.org/10.1109/TVCG.2018.2865240}
\showDOI{\tempurl}


\bibitem[\protect\citeauthoryear{Numbers}{Numbers}{2019}]%
        {thenumbers}
\bibfield{author}{\bibinfo{person}{The Numbers}.} \bibinfo{year}{2019 (accessed
  December 5, 2019)}\natexlab{}.
\newblock \bibinfo{title}{All Time Worldwide Box Office}.
\newblock
\newblock
\urldef\tempurl%
\url{https://www.the-numbers.com/box-office-records/worldwide/all-movies/cumulative/all-time}
\showURL{%
\tempurl}


\bibitem[\protect\citeauthoryear{Peck, Ayuso, and El-Etr}{Peck
  et~al\mbox{.}}{2019}]%
        {peck2019data}
\bibfield{author}{\bibinfo{person}{Evan~M Peck}, \bibinfo{person}{Sofia~E
  Ayuso}, {and} \bibinfo{person}{Omar El-Etr}.}
  \bibinfo{year}{2019}\natexlab{}.
\newblock \showarticletitle{Data is personal: Attitudes and Perceptions of Data
  Visualization in Rural Pennsylvania}. In
  \bibinfo{booktitle}{\emph{Proceedings of the 2019 CHI Conference on Human
  Factors in Computing Systems}}. \bibinfo{publisher}{Association for Computing
  Machinery}, \bibinfo{address}{New York, NY, USA}, \bibinfo{pages}{1--12}.
\newblock
\urldef\tempurl%
\url{https://doi.org/10.1145/3290605.3300474}
\showDOI{\tempurl}


\bibitem[\protect\citeauthoryear{Peer, Brandimarte, Samat, and Acquisti}{Peer
  et~al\mbox{.}}{2017}]%
        {peer2017beyond}
\bibfield{author}{\bibinfo{person}{Eyal Peer}, \bibinfo{person}{Laura
  Brandimarte}, \bibinfo{person}{Sonam Samat}, {and}
  \bibinfo{person}{Alessandro Acquisti}.} \bibinfo{year}{2017}\natexlab{}.
\newblock \showarticletitle{Beyond the Turk: Alternative platforms for
  crowdsourcing behavioral research}.
\newblock \bibinfo{journal}{\emph{Journal of Experimental Social Psychology}}
  \bibinfo{volume}{70} (\bibinfo{year}{2017}), \bibinfo{pages}{153 -- 163}.
\newblock
\showISSN{0022-1031}
\urldef\tempurl%
\url{https://doi.org/10.1016/j.jesp.2017.01.006}
\showDOI{\tempurl}


\bibitem[\protect\citeauthoryear{Qian, Sun, Cui, Lou, Zhang, and Zhang}{Qian
  et~al\mbox{.}}{2020}]%
        {retrievethenadapt}
\bibfield{author}{\bibinfo{person}{Chunyao Qian}, \bibinfo{person}{Shizhao
  Sun}, \bibinfo{person}{Weiwei Cui}, \bibinfo{person}{Jian-Guang Lou},
  \bibinfo{person}{Haidong Zhang}, {and} \bibinfo{person}{Dongmei Zhang}.}
  \bibinfo{year}{2020}\natexlab{}.
\newblock \bibinfo{title}{Retrieve-Then-Adapt: Example-based Automatic
  Generation for Proportion-related Infographics}.
\newblock
\newblock
\showeprint[arxiv]{2008.01177}~[cs.HC]


\bibitem[\protect\citeauthoryear{Shank}{Shank}{2013}]%
        {shank2013computers}
\bibfield{author}{\bibinfo{person}{Daniel~B. Shank}.}
  \bibinfo{year}{2013}\natexlab{}.
\newblock \showarticletitle{Are Computers Good or Bad for Business? How
  Mediated Customer-Computer Interaction Alters Emotions, Impressions, and
  Patronage toward Organizations}.
\newblock \bibinfo{journal}{\emph{Computers in Human Behavior}}
  \bibinfo{volume}{29}, \bibinfo{number}{3} (\bibinfo{date}{May}
  \bibinfo{year}{2013}), \bibinfo{pages}{715–725}.
\newblock
\showISSN{0747-5632}
\urldef\tempurl%
\url{https://doi.org/10.1016/j.chb.2012.11.006}
\showDOI{\tempurl}


\bibitem[\protect\citeauthoryear{Shi, Xu, Sun, Shi, and Cao}{Shi
  et~al\mbox{.}}{2020}]%
        {calliope_2020}
\bibfield{author}{\bibinfo{person}{Danqing Shi}, \bibinfo{person}{Xinyue Xu},
  \bibinfo{person}{Fuling Sun}, \bibinfo{person}{Yang Shi}, {and}
  \bibinfo{person}{Nan Cao}.} \bibinfo{year}{2020}\natexlab{}.
\newblock \bibinfo{title}{Calliope: Automatic Visual Data Story Generation from
  a Spreadsheet}.
\newblock , \bibinfo{numpages}{1}~pages.
\newblock
\showISSN{2160-9306}
\urldef\tempurl%
\url{https://doi.org/10.1109/tvcg.2020.3030403}
\showDOI{\tempurl}


\bibitem[\protect\citeauthoryear{Spotify}{Spotify}{2020}]%
        {spotify}
\bibfield{author}{\bibinfo{person}{Spotify}.} \bibinfo{year}{2020}\natexlab{}.
\newblock \bibinfo{title}{Spotify}.
\newblock \bibinfo{howpublished}{\url{https://www.spotify.com/}}.
\newblock
\newblock
\shownote{Accessed on (2020/01/09).}


\bibitem[\protect\citeauthoryear{{Srinivasan}, {Drucker}, {Endert}, and
  {Stasko}}{{Srinivasan} et~al\mbox{.}}{2019}]%
        {srinivasan2018augmenting}
\bibfield{author}{\bibinfo{person}{A. {Srinivasan}}, \bibinfo{person}{S.~M.
  {Drucker}}, \bibinfo{person}{A. {Endert}}, {and} \bibinfo{person}{J.
  {Stasko}}.} \bibinfo{year}{2019}\natexlab{}.
\newblock \showarticletitle{Augmenting Visualizations with Interactive Data
  Facts to Facilitate Interpretation and Communication}.
\newblock \bibinfo{journal}{\emph{IEEE Transactions on Visualization and
  Computer Graphics}} \bibinfo{volume}{25}, \bibinfo{number}{1}
  (\bibinfo{year}{2019}), \bibinfo{pages}{672--681}.
\newblock
\urldef\tempurl%
\url{https://doi.org/10.1109/TVCG.2018.2865145}
\showDOI{\tempurl}


\bibitem[\protect\citeauthoryear{Tableau}{Tableau}{2021}]%
        {tableauexplain}
\bibfield{author}{\bibinfo{person}{Tableau}.} \bibinfo{year}{2021 (accessed
  January 5, 2021)}\natexlab{}.
\newblock \bibinfo{title}{Explain Data}.
\newblock
\newblock
\urldef\tempurl%
\url{https://www.tableau.com/products/new-features/explain-data}
\showURL{%
\tempurl}


\bibitem[\protect\citeauthoryear{Vaccaro, Karahalios, Mulligan, Kluttz, and
  Hirsch}{Vaccaro et~al\mbox{.}}{2019}]%
        {contestability}
\bibfield{author}{\bibinfo{person}{Kristen Vaccaro}, \bibinfo{person}{Karrie
  Karahalios}, \bibinfo{person}{Deirdre~K. Mulligan}, \bibinfo{person}{Daniel
  Kluttz}, {and} \bibinfo{person}{Tad Hirsch}.}
  \bibinfo{year}{2019}\natexlab{}.
\newblock \showarticletitle{Contestability in Algorithmic Systems}. In
  \bibinfo{booktitle}{\emph{Conference Companion Publication of the 2019 on
  Computer Supported Cooperative Work and Social Computing}}.
  \bibinfo{publisher}{Association for Computing Machinery},
  \bibinfo{address}{New York, NY, USA}, \bibinfo{pages}{523–527}.
\newblock
\showISBNx{9781450366922}
\urldef\tempurl%
\url{https://doi.org/10.1145/3311957.3359435}
\showDOI{\tempurl}


\bibitem[\protect\citeauthoryear{Vartak, Rahman, Madden, Parameswaran, and
  Polyzotis}{Vartak et~al\mbox{.}}{2015}]%
        {vartak2015s}
\bibfield{author}{\bibinfo{person}{Manasi Vartak}, \bibinfo{person}{Sajjadur
  Rahman}, \bibinfo{person}{Samuel Madden}, \bibinfo{person}{Aditya
  Parameswaran}, {and} \bibinfo{person}{Neoklis Polyzotis}.}
  \bibinfo{year}{2015}\natexlab{}.
\newblock \showarticletitle{SeeDB: Efficient Data-Driven Visualization
  Recommendations to Support Visual Analytics}.
\newblock \bibinfo{journal}{\emph{Proc. VLDB Endow.}} \bibinfo{volume}{8},
  \bibinfo{number}{13} (\bibinfo{date}{Sept.} \bibinfo{year}{2015}),
  \bibinfo{pages}{2182–2193}.
\newblock
\showISSN{2150-8097}
\urldef\tempurl%
\url{https://doi.org/10.14778/2831360.2831371}
\showDOI{\tempurl}


\bibitem[\protect\citeauthoryear{Victor, De~Cock, and Cornelis}{Victor
  et~al\mbox{.}}{2011}]%
        {victor2011trust}
\bibfield{author}{\bibinfo{person}{Patricia Victor}, \bibinfo{person}{Martine
  De~Cock}, {and} \bibinfo{person}{Chris Cornelis}.}
  \bibinfo{year}{2011}\natexlab{}.
\newblock \showarticletitle{Trust and recommendations}.
\newblock In \bibinfo{booktitle}{\emph{Recommender systems handbook}}.
  \bibinfo{publisher}{Springer}, \bibinfo{address}{Boston, MA, USA},
  \bibinfo{pages}{645--675}.
\newblock
\urldef\tempurl%
\url{https://doi.org/10.1007/978-0-387-85820-3_20}
\showDOI{\tempurl}


\bibitem[\protect\citeauthoryear{Wongsuphasawat, Moritz, Anand, Mackinlay,
  Howe, and Heer}{Wongsuphasawat et~al\mbox{.}}{2016}]%
        {wongsuphasawat2015voyager}
\bibfield{author}{\bibinfo{person}{Kanit Wongsuphasawat},
  \bibinfo{person}{Dominik Moritz}, \bibinfo{person}{Anushka Anand},
  \bibinfo{person}{Jock Mackinlay}, \bibinfo{person}{Bill Howe}, {and}
  \bibinfo{person}{Jeffrey Heer}.} \bibinfo{year}{2016}\natexlab{}.
\newblock \showarticletitle{Voyager: Exploratory Analysis via Faceted Browsing
  of Visualization Recommendations}.
\newblock \bibinfo{journal}{\emph{IEEE Transactions on Visualization and
  Computer Graphics}} \bibinfo{volume}{22}, \bibinfo{number}{1}
  (\bibinfo{year}{2016}), \bibinfo{pages}{649--658}.
\newblock
\urldef\tempurl%
\url{https://doi.org/10.1109/TVCG.2015.2467191}
\showDOI{\tempurl}


\bibitem[\protect\citeauthoryear{Wongsuphasawat, Qu, Moritz, Chang, Ouk, Anand,
  Mackinlay, Howe, and Heer}{Wongsuphasawat et~al\mbox{.}}{2017}]%
        {wongsuphasawat2017voyager}
\bibfield{author}{\bibinfo{person}{Kanit Wongsuphasawat},
  \bibinfo{person}{Zening Qu}, \bibinfo{person}{Dominik Moritz},
  \bibinfo{person}{Riley Chang}, \bibinfo{person}{Felix Ouk},
  \bibinfo{person}{Anushka Anand}, \bibinfo{person}{Jock Mackinlay},
  \bibinfo{person}{Bill Howe}, {and} \bibinfo{person}{Jeffrey Heer}.}
  \bibinfo{year}{2017}\natexlab{}.
\newblock \showarticletitle{Voyager 2: Augmenting Visual Analysis with Partial
  View Specifications}. In \bibinfo{booktitle}{\emph{Proceedings of the 2017
  CHI Conference on Human Factors in Computing Systems}}.
  \bibinfo{publisher}{Association for Computing Machinery},
  \bibinfo{address}{New York, NY, USA}, \bibinfo{pages}{2648–2659}.
\newblock
\showISBNx{9781450346559}
\urldef\tempurl%
\url{https://doi.org/10.1145/3025453.3025768}
\showDOI{\tempurl}


\bibitem[\protect\citeauthoryear{Yang, Huang, Scholtz, and Arendt}{Yang
  et~al\mbox{.}}{2020}]%
        {YangVisualExplanations}
\bibfield{author}{\bibinfo{person}{Fumeng Yang}, \bibinfo{person}{Zhuanyi
  Huang}, \bibinfo{person}{Jean Scholtz}, {and} \bibinfo{person}{Dustin~L.
  Arendt}.} \bibinfo{year}{2020}\natexlab{}.
\newblock \showarticletitle{How Do Visual Explanations Foster End Users'
  Appropriate Trust in Machine Learning?}. In
  \bibinfo{booktitle}{\emph{Proceedings of the 25th International Conference on
  Intelligent User Interfaces}} (Cagliari, Italy) \emph{(\bibinfo{series}{IUI
  '20})}. \bibinfo{publisher}{Association for Computing Machinery},
  \bibinfo{address}{New York, NY, USA}, \bibinfo{pages}{189–201}.
\newblock
\showISBNx{9781450371186}
\urldef\tempurl%
\url{https://doi.org/10.1145/3377325.3377480}
\showDOI{\tempurl}


\bibitem[\protect\citeauthoryear{Yin, Wortman~Vaughan, and Wallach}{Yin
  et~al\mbox{.}}{2019}]%
        {Yin2019accuracytrust}
\bibfield{author}{\bibinfo{person}{Ming Yin}, \bibinfo{person}{Jennifer
  Wortman~Vaughan}, {and} \bibinfo{person}{Hanna Wallach}.}
  \bibinfo{year}{2019}\natexlab{}.
\newblock \showarticletitle{Understanding the Effect of Accuracy on Trust in
  Machine Learning Models}. In \bibinfo{booktitle}{\emph{Proceedings of the
  2019 CHI Conference on Human Factors in Computing Systems}} (Glasgow,
  Scotland Uk) \emph{(\bibinfo{series}{CHI '19})}.
  \bibinfo{publisher}{Association for Computing Machinery},
  \bibinfo{address}{New York, NY, USA}, \bibinfo{pages}{1–12}.
\newblock
\showISBNx{9781450359702}
\urldef\tempurl%
\url{https://doi.org/10.1145/3290605.3300509}
\showDOI{\tempurl}


\end{thebibliography}

\end{document}